\documentclass{article}

\usepackage{arxiv}
\usepackage{comment}
\usepackage[utf8]{inputenc} 
\usepackage[T1]{fontenc}    
\usepackage{hyperref}       
\usepackage{url}            
\usepackage{booktabs}       
\usepackage{amsfonts}       
\usepackage{nicefrac}       
\usepackage{microtype}      
\usepackage{lipsum}
\usepackage{graphicx}
\graphicspath{ {./images/} }
\usepackage{amsmath}
\usepackage{multirow}
\usepackage{hyperref}
\usepackage{xcolor}
\usepackage{subcaption}

\usepackage{booktabs}
\usepackage{tabularx}
\usepackage{siunitx}
\usepackage[table]{xcolor}
\usepackage{makecell}

\title{Diffusion-Induced Spatial Attention Overlapping Community Detection}

\author{
 Kosti Koistinen \\
  Aalto University School of Science \\
  Computer Science Department \\
  P.O.Box 11000, 00076 \\
  AALTO, Finland \\
    \texttt{kosti.koistinen@aalto.fi} \\
  \And
 Vesa Kuikka \\
  Aalto University School of Science \\
  Computer Science Department \\
  P.O.Box 11000, 00076 \\
  AALTO, Finland \\
  \texttt{vesa.kuikka@aalto.fi} \\
  \And
    Joni Herttuainen \\
  Aalto University School of Science \\
  Computer Science Department \\
  P.O.Box 11000, 00076 \\
  AALTO, Finland \\
  \texttt{joni.herttuainen@aalto.fi} \\
  \And
    Matthew R. Hendren \\
  Lockheed Martin \\
  USA \\
  \texttt{matthew.r.hendren@lmco.com} \\
   \And
    Brian M. Holt \\
  Lockheed Martin \\
  USA \\
  \texttt{brian.m.holt@lmco.com} \\
  \And
    Kimmo K. Kaski \\
  Aalto University School of Science \\
  Computer Science Department \\
  P.O.Box 11000, 00076 \\
  AALTO, Finland \\
  \texttt{kimmo.kaski@aalto.fi} \\
}

\begin{document}
\maketitle
\begin{abstract}
Detection of overlapping communities 
is essential for modelling networks in which nodes participate simultaneously in multiple structural or functional groups. Existing graph neural network approaches commonly rely on local message passing, which can obscure community boundaries through smoothing and limit the representation of structurally relevant long-range dependencies. We introduce Diffusion-Induced Spatial Attention Community Detection (DISCO), a deep-learning framework that combines a structural prior derived from influence spreading dynamics, sparse multi-head attention, and non-negative community-affiliation learning. The prior identifies candidate interactions beyond immediate graph neighbours and biases attention according to their structural proximity, while a Bernoulli–Poisson edge-reconstruction objective enables overlapping community inference from node attributes and 
structural profiles, or both. Benchmark experiments show that DISCO performs competitively against established graph convolutional and graph attention approaches across different input configurations. To demonstrate its practical applicability, we present a proof-of-concept cybersecurity use case in which changes between community assignments inferred from consecutive communication-network snapshots provide an interpretable anomaly signal. Temporal community similarity identifies structural deviations, while node-level contributions help locate the devices associated with them. DISCO therefore provides both a flexible method for overlapping community detection and a foundation for analysing structural change in dynamic networks.
\end{abstract}


\section{Introduction}
Networks provide a natural and expressive representation of complex systems. In a graph-based model, nodes represent entities such as individuals, information-technology devices, or financial accounts, whereas edges represent interactions or relationships among those entities. A wide range of real-world systems can therefore be modeled as networks, including social relationships, communication infrastructures, financial transactions, organisational and biological systems, and beyond. \cite{barabasi2016network}

A common research problem in network science is community detection, which aims to identify groups or clusters of nodes that are more densely connected to each 
other than 
the remainder of the network \cite{fortunato2010community}. Such communities often correspond to meaningful structural or functional units. In social networks, for example, communities can 
represent friendship groups, shared-interest groups, or professional clusters \cite{mcauley2012learning}. In financial and cybercrime networks, transaction patterns can 
reveal groups of coordinated actors engaged in fraudulent or otherwise anomalous activities \cite{akoglu2015graph}. Community detection is also relevant to cybersecurity and vulnerability analysis. Communication networks and critical infrastructures often contain structural communities corresponding to functional subsystems or organisational units. Identifying these structures--and temporal changes in them--can help analysts understand how attacks, failures, or malicious information may spread through a system and locate components whose compromise could cause large-scale disruption \cite{new1}. More generally, community analysis provides insight into how information, influence, risks, and resources flow through a network. Understanding these flows is essential for characterising 
structural organisation and dynamic behavior of the network \cite{computation12040085,akoglu2015graph}.

Community detection emerged as a distinct research topic in the early 2000s. An 
influential early contribution was the Girvan--Newman algorithm, followed by other foundational approaches, including graph-partitioning methods, spectral techniques based on representations such as the graph Laplacian, and modularity optimisation \cite{newman2018networks}. Modularity provided a quantitative objective for evaluating community partitions and subsequently motivated widely used algorithms such as the Louvain method. 
Together, these developments established much of the methodological foundation for modern community detection.
Most 
community-detection methods produce a disjoint partition in which each node belongs to exactly one community. This assumption is convenient computationally but often unrealistic. For example, individuals in social networks often belong simultaneously to family, workplace, educational, and recreational communities \cite{mcauley2012learning}. Similarly, a server or an IT device may participate in several functional subsystems
\cite{akoglu2015graph}. Overlapping community detection addresses this limitation by flexibly allowing each node to belong to more than one community.

Overlapping community detection methods include clique-percolation methods \cite{palla2005uncovering}, label-propagation techniques \cite{gregory2010finding}, and probabilistic frameworks such as the mixed-membership stochastic block model \cite{airoldi2008mixed}. Clique-based methods can provide intuitive representations of overlapping groups, but their performance may deteriorate in sparse networks. Probabilistic mixed-membership models, in turn, can be computationally expensive when applied to large graphs \cite{new2}. Although there have been extensions to these native models, a common limitation is that many traditional methods cannot incorporate both the node and edge attributes, and are often driven by topology 
\cite{yang2013community}. For example, a node representing an individual may be associated with demographic, behavioral, or ideological characteristics, while an edge may contain information about 
the frequency, duration, or type of interaction. Because node attributes, although often 
noisy, can carry important topological information \cite{koistinen2025importance}, reducing a network to its topology alone may discard information that is essential for accurately characterising its structure. Data scarcity further complicates the overlapping-community problem. Reliable ground-truth community memberships are rarely available, particularly when memberships overlap or evolve over time. The node attributes may also be incomplete or completely absent. Consequently, methods that depend heavily on observed attributes may fail when applied to real-world networks. For example, supervised models that require complete information of attributes, can be difficult to train and evaluate. \cite{su2024comprehensive}

Over the past decade, deep-learning methods have been increasingly applied to community detection to resolve the aforementioned limitations. These approaches are attractive because they can represent nonlinear relationships, use high-dimensional node and edge attributes, and scale to complex and large network structures. Early machine-learning techniques, including agglomerative clustering and nonnegative matrix factorisation, were adapted to network clustering before the widespread adoption of deep neural models \cite{yang2013overlapping, yang2012structure}. More recent approaches include autoencoders, adversarial and generative 
networks, convolutional architectures, and graph-specific neural networks. For a complete taxonomy and review, see \cite{su2024comprehensive}.

Most recently, graph neural networks (GNNs) have become particularly influential because they combine deep learning and graph-structured data representations. Architectures such as graph convolutional networks (GCNs) \cite{kipf2017semi} and graph attention networks (GATs) \cite{velickovic2018graph} have consequently been adapted to overlapping community detection tasks \cite{AOCD,shchur2019overlapping}. These architectures make it possible to combine graph topology with node attributes and to learn latent representations that can be used to infer multiple community memberships. However, existing GNN-based methods continue to face limitations related to local message passing, missing attributes, oversmoothing, scalability, noisy graphs, and the representation of dense overlaps, as discussed in detail in Section 2. \cite{li2018, su2024comprehensive}

In this article, we propose the \emph{Diffusion-Induced Spatial Attention Community Detection} (DISCO), a model which incorporates a graph-diffusion prior into the construction of node inputs and the allocation of attention mechanism. The diffusion process describes how signals spread through the network, which allows the model to capture structural relationships beyond immediate neighbourhoods. Although diffusion-based network representations have previously been applied to overlapping community detection \cite{computation12040085}, we demonstrate how diffusion can be integrated into a graph neural architecture and deep learning for effective and scalable overlapping community detection.

We evaluate DISCO on social-network benchmarks with overlapping ground-truth communities and conduct an extensive ablation study to examine the effects of 
prior and input representations.
In addition to methodological benchmarking, we present a proof-of-concept cybersecurity use case, 
in which we investigate the overlapping community structure and its evolution in a dynamic operational technology (OT) network.
More precisely, we will show that anomalous changes in communication patterns can be reflected in changes in the inferred overlapping community structure, which allows for temporal community similarity to serve as an interpretable network-level anomaly signal. However, the purpose of this application 
is not to propose a comprehensive anomaly-detection method but to illustrate the practical potential of DISCO beyond standard benchmark datasets.

The remainder of this paper is organised as follows. Section~2 reviews modern learning-based approaches to the detection of overlapping communities and identifies the research gap addressed by DISCO. Section~3 introduces the proposed model and its principal variants. Section~4 describes the data and experimental setup, and Section~5 presents the benchmark results and the ablation study. Section~6 summarises the results of the proof-of-concept cybersecurity use case, and Section~7 concludes the article and outlines directions for future work.

\section{Related Work}

Here we review the latest works on learning-based approaches to overlapping community detection, with particular emphasis on graph neural networks and attention-based architectures. There are a variety of methods for community detection, but because we focus on overlapping memberships, we will only consider methods that explicitly allow nodes to belong to multiple communities. We investigate 
the principal advantages and limitations of these methods and motivate the need for an approach that captures node attributes, local connectivity, and graph-wide diffusion structure. Rather than providing an exhaustive survey of existing models, we focus on the architectural limitations of model families.

\subsection{Foundational Learning-Based Methods}
Two influential pre-GNN approaches are BigCLAM and CESNA \cite{yang2013overlapping,yang2013community} 
of which the former represents overlapping community affiliations using non-negative latent factors, while the latter 
extends this general framework using node attributes. This allows community membership to be inferred from 
the network structure and observed features. Both approaches are probabilistic, and although these models are well suited to represent multiple memberships, their linear structure can limit their ability to find complex relationships among topology, attributes, and community affiliations \cite{he2022boosting, 3060832.3060936}.

Graph-embedding methods provide an important bridge between traditional structural analysis and later deep learning approaches to community detection. Methods such as LINE \cite{tang2015line} and node2vec \cite{grover2016node2vec} learn low-dimensional node representations by optimising objectives that preserve local neighborhoods and higher-order proximity. These embeddings can then be used as inputs for community detection (see, e.g. \cite{GOYAL201878}, and references therein). In this sense, graph embeddings move beyond graph measures such as degree, shortest-path distance, or common-neighbor counts by learning task-relevant structural representations directly from the network. Although these approaches are not in 
focus of the present study, these approaches mark a broader shift from hand-crafted structural measures toward learned network representations.

\subsection{Graph Convolutional Approaches}

Graph convolutional networks (GCNs) are among the most widely used GNN architectures. Their message-passing layers aggregate information from neighboring nodes, effectively smoothing node representations over the graph. This process is often interpreted as a form of low-pass filtering. Although smoothing can improve representation learning in homophilous networks, where connected nodes tend to have similar attributes, repeated aggregation may blur meaningful discontinuities between communities. This may cause node representations to become indistinguishable. This phenomenon is commonly described as oversmoothing \cite{li2018, oono2020graph}. GCNs may also suffer from underreaching, in which information from structurally relevant but distant nodes cannot be represented within a limited number of message-passing layers \cite{alon2021bottleneck}. These issues partly arise because GCNs were initially developed for tasks such as node classification and link prediction rather than community detection \cite{su2024comprehensive}. Such tasks 
often benefit from local proximity information, whereas community detection may require preserving broader structural patterns and group boundaries.
Because GCN aggregation is restricted to local neighborhoods, feature similarity can dominate the learned representation, making it 
particularly effective in homophilous networks. However, topology may be underrepresented when the node features are highly informative, while the model may perform poorly when attributes are incomplete, unreliable, or unavailable \cite{chen2022learning}. One proposed response is to construct pseudo-attributes from positional or structural characteristics such as node degree and centrality \cite{cui2022positional}.

The Neural Overlapping Community Detection (NOCD) is a popular GCN-based model for overlapping community detection \cite{shchur2019overlapping}. The original evaluation compared a two-layer GCN architecture with adjacency-only and attribute-only variants. In several evaluated cases, NOCD outperformed the compared baseline methods. Notably, the attribute-only model performed better than the topology-based GCN model. This result again supports the claim that the convolutional architecture may not always exploit topology effectively, and thus rely disproportionately on node attributes. A further limitation is that NOCD, like many GNN-based methods, does not directly optimise a conventional community-quality objective such as modularity, but instead optimises by first-order similarity: Connected nodes are assumed to have similar community assignments. This can be misleading in complex networks that contain noisy edges between different communities. The authors in \cite{yuan2023overlapping} argue that NOCD is limited by this objective and propose optimising Markov stability, which more directly evaluates the persistence and quality of the detected community structure. On the other hand, yet another limitation in the representation-learning objective is that it can assign nodes from structurally separate regions to the same community when their attributes are sufficiently similar \cite{new3}. These limitations indicate that effective overlapping community detection requires a balanced integration of both node attributes and graph topology.

\subsection{Graph Attention Approaches}

Graph attention networks (GATs) constitute another modern branch of graph deep learning. Attention mechanisms that 
are widely used in modern natural-language processing, learn the correlations between elements. In a graph, an attention mechanism estimates the relevance of neighboring nodes when constructing the representation of a focal node. Standard GATs generally mask attention to observed neighbors, which 
enables the model to assign different weights to local relationships, unlike conventional GCNs that 
apply a more uniform convolutional aggregation \cite{velickovic2018graph}. Nevertheless, graph attention remains local: via masking, a node can directly attend only to nodes connected by an edge.

In \cite{AOCD}, the authors incorporate a GAT into the NOCD framework by replacing its two-layer GCN with Attention. The resulting model was evaluated on the same datasets as NOCD and achieved better performance in most reported cases with similar complexity. The authors also reported rapid convergence, which indicates that attention-based aggregation can offer computational as well as predictive advantages. Despite these benefits, local attention does not fully resolve the limited receptive field of message-passing networks. Ideally, a node representation should be able to find relevant interactions from across the graph rather than only from its immediate neighborhood. Full pairwise self-attention would allow every node to attend to every other node, but its computational complexity grows as $(O(N^2))$ \cite{shirzad2023exphormer}. Full attention may also increase the risk of overfitting by introducing a large number of potentially irrelevant interactions, as attention poses risk of spurious correlations \cite{koistinen2026spatiotemporalattentiongraphneural}. Furthermore, multi-head-attention and other aggregating mechanisms pose a risk for redundancy and inefficiency \cite{ye2023sparse}. 

Scalable sparse-attention mechanisms have become 
popular in response to the limitations of local message passing. Recent graph-transformer architectures aim to move beyond the local message-passing paradigm via allowing nodes to access broader structural context through attention and structural encodings. This represents a shift toward models that can capture longer-range dependencies in the graph. However, many graph-transformer designs are adapted from transformer architectures originally developed for sequential language data, and therefore require additional positional or structural encodings to properly reflect graph topology \cite{ying2021graphormer}. In \cite{shirzad2023exphormer}, this problem is addressed through a sparse attention scheme in which randomly added edges allow nodes to attend to selected nodes in distant regions of the network. This construction exposes the model to more global connectivity patterns while retaining approximately $O(N+E)$ complexity, rather than the $O(N^2)$ complexity of full attention. However, randomly rewired connections do not necessarily correspond to meaningful structural relationships \cite{koistinen2025importance}.

\subsection{Diffusion-Guided Graph Learning}

The present study builds on the principle of extending attention beyond observed local edges. Rather than adding random connections, we construct additional relationships via probabilistic influence spreading model, introduced in \cite{kuikka2018influence}. A diffusion process provides a probabilistic measure of how signals, information, or influence propagate through the network. It can therefore identify structurally relevant relationships between nodes that are not immediate neighbors \cite{kuikka2024detailed}. Graph diffusion is already a widely studied topic in community detection; see \cite{GAO2025128396}; However, only a limited number of studies have explicitly combined diffusion-based graph dynamics with GNN-based community detection. Closely related examples include CDMG, which optimises a Markov-stability objective with a GCN \cite{yuan2023overlapping}, and GDCL, which uses graph diffusion and graph convolutional contrastive learning for community detection \cite{zhang2026graph}. This suggests that the interaction between graph diffusion, neural attention, and overlapping community detection remains comparatively underexplored.

\subsection{Research Gap}

Existing modern learning-based methods either directly model overlapping memberships 
or learn nonlinear node representations, but 
often do not fully resolve how to include structurally meaningful nonlocal information. A natural response is to extend the interaction space of the model. However, unrestricted global attention is computationally expensive and may introduce irrelevant interactions, while randomly added sparse connections do not necessarily reflect meaningful graph structure. For detection of overlapping communities, nonlocal interactions should therefore be selected according to a structural principle rather than being introduced arbitrarily. These observations motivate our Diffusion-Induced Spatial Attention Community Detection (DISCO) model. The proposed model uses diffusion-derived relationships as a structural prior for sparse spatial attention. Diffusion extends the interaction space beyond the immediate graph neighbors, while avoiding the cost and potential noise of unrestricted global attention. Unlike randomly introduced nonlocal edges, diffusion-induced connections are based on a graph spreading process and therefore provide a principled way to identify structurally relevant candidate interactions. By combining this prior-guided attention with flexible input representations and nonnegative community-affiliation learning, DISCO is designed to capture both conventional boundary overlaps and dense overlapping memberships whose identification depends on local topology, node attributes, and higher-order graph structure.

Beyond identifying communities in a static graph, overlapping representations can also provide a basis for analysing structural changes over time. 
Although dynamic community detection has been studied extensively~\cite{rossetti2018survey}, and graph-based anomaly detection is a well-established field of research~\cite{akoglu2015graph}, temporal community detection in OT environments remain underexplored.
One related exception is the work in ~\cite{safdari2024anomalydetection}, where the authors used the community structure of the networks as a signal for anomaly detection in research. 
For network intrusion detection, static community-based methods have been explored~\cite{ding2012intrusion}, but utilising a change-based signal in a dynamic setting has received limited attention. Some recent theoretical work on measuring the temporal changes in community structure exists, such as \cite{zhong2025quantifying}, in which the authors introduce modified metrics based on $\mathrm{ONMI}$ to quantify the changes but do not evaluate the methodology in an anomaly or intrusion detection context.
Motivated by this, we present 
an experiment utilising OT network data to demonstrate how our approach can be used to detect cyberattacks and malfunctions in communication networks.

\section{Model}

Here we propose a Diffusion-Induced Spatial Attention Community Detection model (\textbf{DISCO}), the goal of which 
is to learn the overlapping community structure in graphs by combining data-driven representation learning with structurally-informed priors that are derived from the Influence Spreading Model (ISM) defined in \cite{kuikka2018influence}. We consider a graph $\mathcal{G} = (\mathcal{V}, \mathcal{E})$ with $N = |\mathcal{V}|$ nodes and edges $\mathcal{E} \subseteq \mathcal{V} \times \mathcal{V}$. 
Each node $i \in \mathcal{V}$ may optionally be associated with a feature vector $x_i \in \mathbb{R}^D$, forming a feature matrix $X \in \mathbb{R}^{N \times D}$, where $D$ denotes the number of features. The model overview and its modules are illustrated in Figure~1. Next, we describe each in detail.

\begin{figure}
    \centering
    \includegraphics[width=0.98\linewidth]{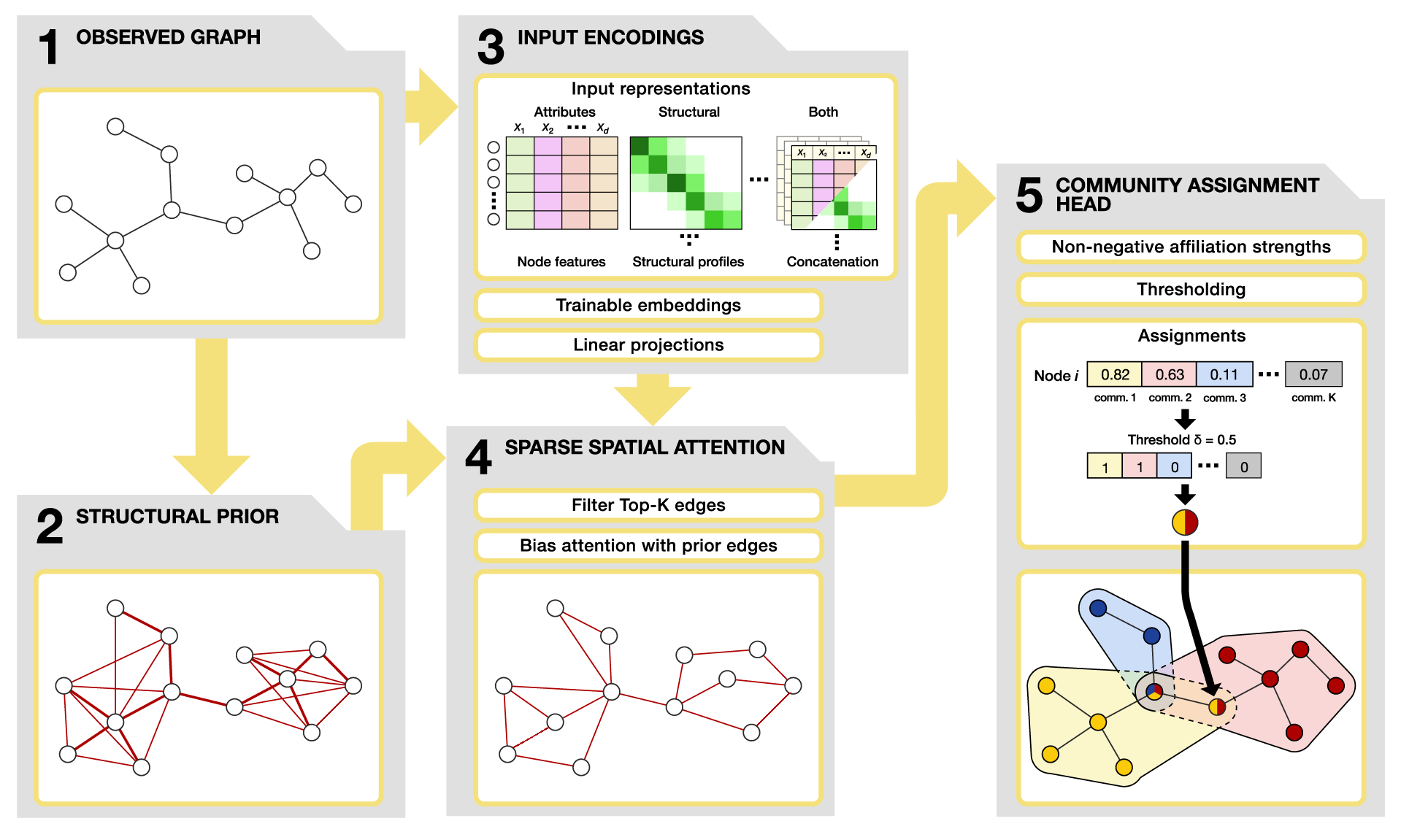}
    \caption{Overview of the DISCO architecture, combining diffusion-derived structural priors, flexible node representations, spatial multi-head attention, and overlapping community assignment.
}
    \label{fig:placeholder}
\end{figure}

\subsection{Architecture}
\subsubsection*{Structural Prior Module}
In addition to the observed graph \(\mathcal{G}\), DISCO incorporates a structural prior graph defined as 

\[
\mathcal{G}_{\mathrm{prior}} = \left(\mathcal{V}, \mathcal{E}_{\mathrm{prior}}, B\right),
\]

where \(B \in \mathbb{R}_{\geq 0}^{N \times N}\) is a nonnegative structural-prior matrix. Each entry \(b_{ij}\) represents the structural prior score associated with the ordered node pair \((i,j)\), specifically implemented as the probabilistic influence spreading value produced by the ISM \cite{kuikka2018influence}. The purpose of the prior graph is to expose the subsequent attention mechanism to structurally significant interactions that may not be represented as observed edges in \(\mathcal{E}\). Unlike standard message-passing architectures, which limit aggregation to immediate neighbors, the prior allows a node to attend to selected nonlocal nodes whose relationships are supported by the global diffusion structure of the graph.

In this work, the matrix \(B\) is derived from spreading process operating on the observed graph. To govern this process, we assign a uniform link weight \(w\) to all observed edges. It is essential to differentiate between two distinct weight concepts in our framework: first, the observed initial link weight \(w\), which dictates the local transition probabilities of the diffusion process; and second, the resulting spreading probability \(b_{ij}\), which is generated by the ISM and utilised directly by the spatial attention block as an additive structural bias in the attention score computation.

Intuitively, \(b_{ij}\) measures the probability that a diffusion process initiated at node \(i\) successfully reaches node \(j\). Thus, \(b_{ij}\) is not merely a function of direct adjacency; it also reflects multi-hop connectivity, shared neighborhoods, and alternative paths within the network. This is particularly valuable for overlapping community detection, as nodes belonging to the same community do not need to be directly connected, especially in sparse or noisy graphs. The ISM computes these pairwise spreading probabilities by allowing influence to propagate freely along graph links without conservation of influence. To ensure that the prior effectively distinguishes between strongly and weakly related node pairs, we assign a carefully balanced uniform link weight of \(w = 0.05\) to the observed graph edges. Previous work has shown that small weights can preserve informative network structure in relatively small social networks \cite{koistinen2025importance}. The choice of $w$ is particularly important because excessively large weights cause the spreading probabilities to approach one, reducing the contrast between node pairs, whereas very small weights cause the probabilities to approach zero and limit propagation beyond local neighbourhoods. The choice is further discussed in Appendix A.

Although the ISM framework can also be used to compute overlapping communities directly~\cite{computation12040085}, DISCO uses its resulting spreading probabilities \(b_{ij}\) as a continuous structural prior for sparse attention. We emphasise that the ISM is just one possible choice of prior; our framework does not rely exclusively on this specific construction. For example, the binary adjacency matrix of \(\mathcal{G}\) could alternatively be used by setting \(b_{ij}=1\) when there is an edge between nodes \(i\) and \(j\), and \(b_{ij}=0\) otherwise. 
For a detailed description of ISM and its parameters, see \cite{kuikka2018influence}.

\subsubsection*{Input Representation Module}
The Input Representation Module constructs initial node representations $H^{(0)} \in \mathbb{R}^{N \times d}$ using one of several input modes:
\[
H^{(0)} =
\begin{cases}
\phi_X(X), & \text{node-attribute input},\\
\phi_R(R), & \text{structural-profile (adjacency) input},\\
\phi_{XR}([X \| R]), & \text{concatenated attribute--structural input},
\end{cases}
\]

where $X \in \mathbb{R}^{N \times D}$ is the node feature matrix, and $R$ is a row-wise structural-profile matrix. The structural profile can be derived either from the binary adjacency matrix of the observed graph or from 
a prior. The operator $[X \| R]$ denotes feature concatenation. The functions $\phi_X$, $\phi_R$, and $\phi_{XR}$ are learnable projections in the embedding dimension of the model $d$. In the node-attribute input mode, the representations are obtained solely from the projected node attributes. In the structural-profile input mode,
the representations are obtained solely from topologically derived 
structural profiles. In the concatenated attribute--structural input mode, the model combines node attributes with structural profiles, allowing it to use both feature-derived and topology-driven
information. 

\subsubsection*{Spatial Multi-Head Attention Block}

The core of DISCO is the spatial multi-head attention mechanism that propagates information between nodes and uses structural priors. The node representations produced by the input representation module are used as input to the attention block. Given these initial representations $H^{(0)}$, we compute queries, keys, and values:
\[
Q = H^{(0)} W_Q, \quad
K = H^{(0)} W_K, \quad
V = H^{(0)} W_V,
\]
where $W_Q, W_K, W_V \in \mathbb{R}^{d \times d}$ are learnable parameters. The embedding dimension is divided across $h$ attention heads $r \in [1,...,h]$, each of dimension $d_h = d / h$. The projected matrices are split across heads as $Q = [Q^{(1)}, \ldots, Q^{(h)}]$, $K = [K^{(1)}, \ldots, K^{(h)}]$, and $V = [V^{(1)}, \ldots, V^{(h)}]$, where $Q^{(r)}, K^{(r)}, V^{(r)} \in \mathbb{R}^{N \times d_h}$. For each attention head, the interaction scores between node pairs can be written as:
\[
S^{(r)} = \frac{Q^{(r)} {K^{(r)}}^\top}{\sqrt{d_h} \cdot \tau} + \lambda B,
\]
where $S^{(r)} \in \mathbb{R}^{N \times N}$ is the matrix of attention logits for head $r$, $B \in \mathbb{R}^{N \times N}$ is the structural prior matrix with entries $b_{ij}$, $\tau$ is a learnable temperature parameter, and $\lambda$ is a learnable scaling factor that controls 
the influence of the prior. The first term corresponds to scaled dot-product attention logits, while the second term introduces an additive bias derived from the prior graph. We adopt an additive formulation rather than a multiplicative one so that the prior enters the attention mechanism as a structural bias, following graph Transformer architectures that encode spatial or edge information as attention biases~\cite{ying2021graphormer}. This avoids directly gating the learned dot-product scores by the prior, which could overly amplify or suppress interactions when the prior is noisy or imperfectly calibrated.

Disco can be used in sparse and dense attention settings. When sparse attention is used, the score matrix $S^{(r)}$ is masked so 
that only the entries corresponding to $(i,j) \in \mathcal{E}_{\text{prior}}$ are retained, and all other entries are set to $-\infty$ before normalisation. Otherwise, attention is computed over all node pairs, with the prior acting solely as an additive bias on the attention logits. Attention weights are obtained by applying a row-wise softmax activation function:
\[
A^{(r)} = \text{softmax}(S^{(r)}),
\]
where $A^{(r)} \in \mathbb{R}^{N \times N}$ and each row sums to one. For each attention head, the value representations are then aggregated as:
\[
Z^{(r)} = A^{(r)} V^{(r)}.
\]
The outputs of all attention heads are concatenated along the feature dimension and normalised using layer normalisation:
\[
H' = \mathrm{LayerNorm}\left(\mathrm{Concat}\left(Z^{(1)}, Z^{(2)}, \ldots, Z^{(h)}\right)\right).
\]
The selected attention mechanism enables the model to combine learned similarity (captured by the dot-product $Q^{(r)}K^{(r) \ \top}$) with structural proximity (captured by $B$). The prior thus acts as a soft inductive bias rather than a hard constraint, allowing the model to adaptively balance structural information and learned representations.

Our formulation of attention is closely related to attention-based graph neural networks such as Graph Attention Network (GAT) \cite{velickovic2018graph}, but differs in a key aspect. GAT employs additive attention based on a learned compatibility function over concatenated node representations, whereas DISCO uses scaled dot-product attention, as in Transformer architectures, and augmented with an additive structural bias. This design yields a more expressive interaction mechanism: dot-product attention captures richer pairwise relationships than additive attention, and allows the model to operate over both observed and prior-induced connectivity structures \cite{ying2021graphormer}.

\subsubsection*{Community Assignment Head}
The final node representations $H'$ are mapped to community affiliations. We use ReLU activation function to allow unbounded non-negative memberships, which aligns with Bernoulli-Poisson factorisation assumptions, as in \cite{AOCD, shchur2019overlapping}:
\[
F = \text{ReLU}(H' W_c),
\]
where $W_c \in \mathbb{R}^{d \times K}$ is a learnable projection matrix and $F \in \mathbb{R}_+^{N \times K}$ is the non-negative community affiliation matrix. Each row $F_i$ represents the strength of membership of node $i$ in 
$K$ communities that 
allows nodes to belong to multiple communities simultaneously. The learned affiliations are later used to define a probabilistic model over edges, where nodes with higher community overlap are more likely to be connected.

\subsection{Training Objective}

The goal of DISCO is to learn model parameters \(\theta\) that produce a
non-negative community-affiliation matrix
\(F_\theta \in \mathbb{R}_+^{N \times K}\). Each row
\(F_i \in \mathbb{R}_+^K\) represents the strengths of node \(i\)'s
memberships in \(K\) communities. We adopt the Bernoulli--Poisson (BP)
formulation used in NOCD \cite{shchur2019overlapping}, in which the
probability of a positive relation between nodes \(i\) and \(j\) is

\[
p_{ij} = 1 - \exp(-F_i^\top F_j).
\]

Thus, node pairs with greater community-affiliation overlap are assigned a
higher probability of being connected.

The structural prior used by the attention mechanism and the node pairs used in the reconstruction objective are conceptually distinct. Let \(\mathcal{E}_{\mathrm{rec}}\) denote the positive pairs used for training. In the standard setting without a prior, \(\mathcal{E}_{\mathrm{rec}} = \mathcal{E}\), where \(\mathcal{E}\) is the
set of observed graph edges. If structural prior is provided,
\(\mathcal{E}_{\mathrm{rec}} = \mathcal{E}_{\mathrm{diff}}\), where \(\mathcal{E}_{\mathrm{diff}}\) contains the undirected node pairs retained after possibly thresholding the diffusion prior. Consequently, diffusion-derived non-neighbour pairs may be treated as positive reconstruction pairs. Let \(\mathcal{E}_{\mathrm{rec}}^-\) denote randomly sampled node pairs that are not contained in \(\mathcal{E}_{\mathrm{rec}}\). The BP loss becomes

\[
\mathcal{L}_{\text{BP}} =
\frac{1}{|\mathcal{E}_{\mathrm{rec}}|}
\sum_{(i,j) \in \mathcal{E}_{\mathrm{rec}}}
\left[-\log\left(1 - \exp(-F_i^\top F_j)\right)\right]
+
\frac{1}{|\mathcal{E}_{\mathrm{rec}}^-|}
\sum_{(i,j) \in \mathcal{E}_{\mathrm{rec}}^-}
F_i^\top F_j.
\]

The first term increases the affiliation overlap of positive pairs, whereas the second term penalises overlap between sampled negative pairs.

To encourage sparse and interpretable community assignments, we add an
\(\ell_1\) regularisation term:

\[
\mathcal{L}_{\text{sparsity}} =
\lambda_{\text{reg}} \|F\|_1
= \lambda_{\text{reg}} \sum_{i=1}^{N} \sum_{k=1}^{K} |F_{ik}|,
\]

where \(\lambda_{\text{reg}} > 0\) controls the sparsity penalty. The final
training objective is

\[
\mathcal{L} = \mathcal{L}_{\text{BP}} + \mathcal{L}_{\text{sparsity}}.
\]

In addition, weight decay is applied to the trainable model parameters.
Weight decay regularises the neural-network parameters, whereas the
\(\ell_1\) term acts directly on the learned community affiliations.

\subsection{Computational Complexity}

A key distinction between DISCO and standard graph neural networks such as GCN and GAT is the scope of node interactions. Traditional methods restrict message passing to immediate neighbors in each layer, whereas DISCO uses the structural prior to define an alternative set of candidate edges. This allows information to propagate through prior-induced connections, which may include long-range node interactions within a single attention layer. This increased expressiveness comes with a computational trade-off. In the dense setting, where attention is computed over all node pairs, the number of candidate interactions scales as $\mathcal{O}(N^2)$.
In the sparse setting, attention is computed only over the retained prior edge set \(\mathcal{E}_{\text{prior}}\), giving rise to complexity $\mathcal{O}(|\mathcal{E}_{\text{prior}}|)$.
The size of \(\mathcal{E}_{\text{prior}}\) depends on the thresholding strategy used to construct the prior graph. The structural prior may add long-range candidate edges, remove weak observed edges, or both (see Appendix B for analysis). Therefore, \(\mathcal{E}_{\text{prior}}\) is not necessarily larger than the observed edge set \(\mathcal{E}\). When the retained prior graph remains sparse, the computational cost satisfies
\[
\mathcal{O}(|\mathcal{E}_{\text{prior}}|) = \mathcal{O}(r|\mathcal{E}|)\ll \mathcal{O}(N^2),
\]

where $r$ is the edge ratio of prior and original sparse graph $\mathcal{E}$. Thus, DISCO provides a practical compromise between the quadratic cost of dense attention and the purely local interactions of standard GAT. By controlling the sparsity of the prior edge set, DISCO can allow richer structural interactions while keeping the attention computation substantially below the dense all-pairs setting.

\section{Data and Graph Construction}

\subsection*{Benchmark Data}
For benchmarking, we use a subset of the Facebook ego-network datasets provided by the Stanford Network Analysis Project (SNAP). The original collection was introduced by McAuley and Leskovec for the study of automatic social-circle discovery \cite{mcauley2012learning}. It contains ten anonymised ego networks collected from Facebook users who participated in a survey. From this collection, we 
select the six ego networks identified by the ego-user IDs 348, 414, 686, 698, 1684, and 1912. These networks have also been used for benchmarking in several previous studies, which enables a direct comparison with other community-detection methods. The main statistics of the selected networks are reported in Table~\ref{tab:dataset_statistics}.

Each dataset represents the local social network of a single Facebook user, referred to as the ego. The remaining nodes, commonly called alters, correspond to the ego's Facebook friends. An undirected edge between two alters indicates that they are also friends with each other, forming a triad. In addition to the graph structure, the datasets provide manually specified social circles, or friend lists, for each ego user. These circles represent groups such as family members, colleagues, classmates, or other socially meaningful sets of friends, and the ground truths are established from these circles. A node may belong to multiple circles simultaneously; for example, a person may be both a colleague and a close friend. The datasets therefore provide overlapping ground-truth community memberships rather than a disjoint partition of the nodes, and this property makes them particularly suitable for evaluating overlapping community detection methods.

The datasets also contain anonymised profile attributes for the anonymised nodes. The original Facebook profile information was collected from multiple categories, including hometowns, birthdays, workplaces or colleagues, educational information, and political and religious affiliations. These attributes are represented as binary features indicating whether a particular profile property applies to a user or not.

\begin{table}[ht]
    \centering
    \caption{Statistics of the Facebook network datasets, where $N$ denotes the number of nodes, $M$ the number of edges, $D$ the number of node attributes, and $C$ the number of ground-truth communities.}
    \label{tab:dataset_statistics}
    \begin{tabular}{lrrrr}
        \toprule
        \textbf{Network} &
        \textbf{$N$} &
        \textbf{$M$} &
        \textbf{$D$} &
        \textbf{$C$} \\
        \midrule
        348  & 224 & 3.2K  & 161 & 14 \\
        414  & 150 & 1.7K  & 105 & 7  \\
        686  & 168 & 1.6K  & 63 & 14 \\
        698  & 61  & 0.3K   & 48 & 12 \\
        1684 & 786 & 14.0K & 319 & 17 \\
        1912 & 747 & 30.0K & 480 & 46 \\
        \bottomrule
    \end{tabular}
\end{table}

\subsubsection*{Cybersecurity Data}
In order to obtain relevant and reliable data for the cybersecurity experiment, we configured a dedicated OT/IT network testbed emulating an industrial control system (ICS) environment to produce network traffic.
The topology implements a layered, front-to-back data pipeline comprising programmable logic controllers (PLCs), industrial surveillance cameras, and virtual and physical workstations.
A Mac Mini serves as the primary entry point, simulating user interaction with the web interface of a Siemens S7-1200 PLC, as well as communication with a MicroLogix 1100 PLC via Common Industrial Protocol (CIP).
Data flow branches into two parallel processing paths. In the first branch, the Siemens PLC issues Modbus/TCP write commands to the Schneider M221 PLC.
Register values stored within the Schneider PLC are transmitted over Modbus/TCP to a virtualized Linux host, denoted Historian \#1.
In the parallel branch, a second virtualized Historian receives CIP packets from the MicroLogix PLC.
Streams from two Axis network cameras are captured by respective Historian virtual machines (VM) via containerized microservices to parse and store video data into local databases.
Following data ingestion in both branches, each Historian publishes MQTT messages to a dedicated Broker VM, which consolidates the data from both branches.
Finally, three Subscriber VMs consume these messages based on specific topic subscriptions.
All networking scripts are orchestrated using containerized microservices deployed in tandem using Docker Compose, and Ansible is used for declarative provisioning of network states with seamless transition.
An APC AP7932 Switched Rack PDU has been provisioned for remote power-cycling of PLCs.
Additional static device attributes are tabulated in Appendix~\ref{app:device_list}.

Traffic is captured via \texttt{tcpdump} from two network interfaces: (a) a physical network-switch SPAN port, and (b) an hypervisor-internal mirrored switch capturing VM correspondence.
We devised Ansible playbooksto control precisely-timed deviations in network operation exercising anomaly classes such as host dropout, role‑change, unexpected port usage, timing‑drift, and traffic‑volume shifts.
For each anomaly case, generated PCAP files corresponding to each interface are filtered using the \texttt{tshark} utility and merged using \texttt{mergecap}.

\begin{figure}[ht]
    \centering
    \includegraphics[width=1.0\linewidth]{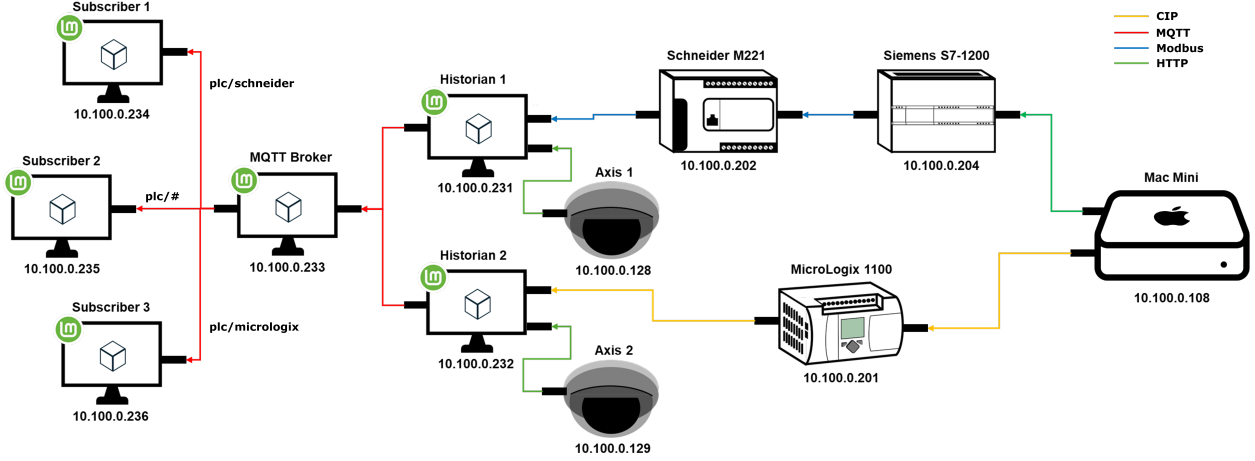}
    \caption{Simulated OT Network Graph.}
    \label{fig:ot_nw}
\end{figure}

The network traffic is divided into consecutive aggregation windows and a weighted graph is constructed for each window.
For each aggregation window, the flow records are converted into a weighted communication graph. Devices are represented as nodes, and an edge is introduced between two devices when communication between them is observed during the corresponding window. The edge weights represent the volume of communication between the associated devices. In the present implementation, the weight of an edge is derived from the sum of bytes observed between the devices within the aggregation window. Alternative definitions, such as packet or flow count, byte volume, or communication frequency, could also be considered depending on the characteristics of the monitoring environment. Finally, we note that the ISM-based structural prior was not used in this experiment. The communication network is very small, which limits the benefit of introducing diffusion-derived nonlocal interactions. It also makes the construction of a meaningful ISM prior very sensitive to the selected diffusion parameters. We therefore use the constructed weighted structure directly as a structural prior. This is consistent with the general DISCO formulation, which does not require the prior to be diffusion-derived and can alternatively operate using weighted adjacency matrix.

As the raw flow records do not necessarily contain suitable semantic node attributes, we construct topology-derived pseudo-attributes for the graph neural network. Specifically, node degree and PageRank are used as node-level features. Degree describes the local connectivity of a device, whereas PageRank provides a measure of its relative structural importance in the communication graph. These features enable the use of node-driven graph neural network architectures even when explicit device attributes are unavailable. Their importance and effect in community detection has been studied in \cite{cui2022positional}.

The choice of aggregation-window length determines the temporal and structural resolution of the resulting graph sequence. Very short windows ($<1$ s) may produce sparse and highly variable graphs. 
In contrast, very long windows ($>10 $ s) may smooth out short-lived changes and produce an excessively 
flat anomaly-score signal. The trade-off is particularly relevant in operational technology environments, where communication is often periodic and deterministic.
We therefore consider aggregation windows of one second, based on heuristic evaluation. 
The window size should be selected empirically for each network to suppress short-term fluctuations while preserving structurally meaningful changes in communication behavior.

\subsubsection*{Cybersecurity Anomaly Scenarios}
The model is evaluated using 10 controlled anomaly scenarios. Each scenario 
is designed to represent changes in the communication structure that could emulate 
a cyberattack.

\begin{enumerate}

\item \textbf{Stale asset.}
A device that normally communicates at regular intervals becomes inactive and stops transmitting traffic. Its disappearance removes expected connections from the network and should alter the inferred community structure.

\item \textbf{Changed device functionality.}
A device stops performing its usual functions and begins generating a different type of network traffic. This scenario changes both its communication partners and its structural role in the network, potentially affecting the surrounding community assignments.

\item \textbf{Fingerprinted Traffic.}
A new communication flow appears between devices that did not previously communicate. The introduction of packets between these hosts creates a new network connection and may change the local topology and the inferred relationships between devices.

\item \textbf{New Port between known hosts.}
Two devices that already communicate begin exchanging traffic through previously unused source or destination ports. Although the hosts remain the same, the new port introduces a different traffic fingerprint and may indicate the use of a new protocol, service, or device function.

\item \textbf{Changed communication frequency.}
The frequency or volume of communication between selected devices is 
modified. Although the devices may remain connected, the altered edge 
weights or temporal interaction pattern may change their structural roles 
and community memberships.

\item \textbf{Missing communication.}
A previously recurring communication relationship between selected devices 
ceases during the anomalous period. This scenario represents, for example, 
a device failure, communication interruption, service disruption, or the 
removal of an expected network flow.

\item \textbf{Pattern Repeat Failure.}
A recurring sequence of related communications is interrupted for one expected repetition. One or more transmissions in the established communication chain are missing, after which the normal pattern resumes, creating a temporary break in the learned temporal dependency.

\item \textbf{Pattern stops repeating.}
A recurring sequence of related communications is interrupted and does not resume during the remainder of the observation period. The permanent loss of the expected repetitions indicates that one stage of the established communication chain has stopped functioning.

\item \textbf{More traffic between known hosts.}
The frequency of packet transmission between devices that already communicate increases significantly. The hosts and communication relationship remain unchanged, but the increased traffic volume produces a stronger edge weight and a denser temporal interaction pattern.

\item \textbf{Less traffic between known hosts.}
The frequency of packet transmission between devices that already communicate decreases significantly. The connection remains present, but its reduced traffic volume weakens the corresponding edge weight and may alter the devices' inferred structural roles or community memberships.

\end{enumerate}

Each experimental sequence contains a 10 hour baseline period, followed by a 1 hour anomalous period and a 1 hour recovery period containing normal traffic. This design makes it possible to examine two properties of the anomaly signal. First, we evaluate whether the temporal $\mathrm{ONMI}$ score changes when the anomaly is introduced. Second, we examine whether the inferred community structure and $\mathrm{ONMI}$ score return toward their baseline levels after normal communication resumes.

\section{Experimental Setup}

\label{sec:onmi_mgh}
\subsection{Evaluation}
\subsubsection*{Overlapping Normalized Mutual Information Score}
To evaluate the agreement between the communities detected by the proposed model and the reference community structure-- i.e., ground truth-- we use the Overlapping Normalized Mutual Information based on the McDaid--Greene--Hurley formulation, denoted as $\mathrm{ONMI}$ \cite{mcdaid2011normalized}. This measure extends normalized mutual information to the case where nodes may belong to multiple communities. Therefore, it is suitable for evaluating overlapping community detection, where the output is a cover rather than a disjoint partition. The value of $\mathrm{ONMI}$ lies in the interval $[0,1]$, where $1$ indicates identical overlapping community covers and values closer to $0$ indicate weaker agreement between the detected and reference covers. In this study, higher $\mathrm{ONMI}$ values are interpreted as better recovery of the reference overlapping community structure. A detailed definition and derivation of the measure are provided in the original work by McDaid et al. \cite{mcdaid2011normalized}.

\subsubsection*{Temporal Community-Similarity Measure}
In addition to benchmarking against ground-truth communities, we use $\mathrm{ONMI}$ to quantify temporal changes in the inferred community structure in demonstrations in Section \ref{sec:anomaly_detection}. Let $(\mathcal{C}_t)$ denote the overlapping community assignment obtained for a binned timestep (t). Temporal stability is measured by comparing consecutive assignments:

\begin{align*}
    s_t = \operatorname{ONMI}(\mathcal{C}_{t-1},\mathcal{C}_t).
\end{align*}

Values close to 1 indicate that the overlapping community structure has remained stable, whereas lower values indicate greater changes in node memberships between consecutive windows. An unusually low value of $s_t$ may therefore identify a time window in which the network departs from its recent structural behavior. Conversely, an unusually high value may indicate a malfunction that causes the network structure to become abnormally static.

Because ordinary fluctuations may also reduce the temporal $\mathrm{ONMI}$ score, individual values should be interpreted relative to the variation observed during a baseline period representing normal network behavior. A detection threshold can be estimated from the empirical distribution of the baseline scores. Possible strategies include selecting a lower percentile of the baseline distribution or defining the threshold as a chosen number of standard deviations below the baseline mean. In the present experiments, we use a $(3\sigma)$ criterion. A window is considered anomalous when its temporal $\mathrm{ONMI}$ score falls below the resulting threshold. The thresholding rule should nevertheless be regarded as environment-dependent, because the normal variability of the community structure differs across networks, aggregation-window lengths, and operating conditions.

\subsection{Configuration}
\subsubsection*{Benchmark}
For evaluating DISCO against others, we use the model and prior parameters shown in Table~\ref{tab:model_parameters}. Most of the parameters are similar to experiments in \cite{shchur2019overlapping,AOCD}. The prior threshold ratio is set to 1.0 for benchmarking, but it could be any. In some cases, shown in Appendix B, the decrease or increase of the ratio changes the outcome and an optimal value depends on the network size and type.
\begin{table}[ht]
    \centering
    \caption{Model and training parameters used in the benchmark experiments. Parameters are grouped according to the corresponding DISCO module or training stage.}
    \label{tab:model_parameters}
    \begin{tabular}{lll}
        \toprule
        \textbf{Parameter} &
        \textbf{Symbol} &
        \textbf{Value} \\
        \midrule

        \multicolumn{3}{l}{\textit{Structural Prior Module}} \\
        Influence Spreading initial edge weight
            & $w$
            & $0.05$ \\
        Maximum path length
            & $L$
            & $100$ \\
        Prior edge ratio
            & $r$
            & $1.0$ \\

        \addlinespace
        \multicolumn{3}{l}{\textit{Input Representation and Spatial Attention}} \\
        Embedding dimension
            & $d$
            & $24$ \\
        Number of attention heads
            & $h$
            & $6$ \\
        Structural-prior bias scale
            & $\lambda$
            & $0.6$ \\
        Dropout probability
            & --
            & $0.1$ \\

        \addlinespace
        \multicolumn{3}{l}{\textit{Community Assignment and Training Objective}} \\
        Community-assignment threshold
            & --
            & $0.5$ \\
        Affiliation sparsity regularisation
            & $\lambda_{\mathrm{reg}}$
            & $10^{-6}$ \\
        Weight decay
            & --
            & $10^{-2}$ \\

        \addlinespace
        \multicolumn{3}{l}{\textit{Training Configuration}} \\
        Learning rate
            & --
            & $10^{-4}$ \\
        Batch size
            & --
            & $10000$ \\
        Maximum training epochs
            & --
            & $5000$ \\
        Early-stopping patience
            & --
            & $500$ epochs \\
        Number of random initialisations
            & --
            & $50$ \\

        \bottomrule
    \end{tabular}
\end{table}

\subsubsection*{Cybersecurity setup}
Training a separate graph neural network for every aggregation window would be computationally infeasible. We therefore use an ensemble of independently trained models. In the current experimental setup, 10 models are trained using 10 randomly selected windows 
from the baseline period, i.e., from the period of normal traffic without anomalies. We use same parameters as in benchmarking. We remind, however, that choosing them is heuristic and environment-dependent. Furthermore, the models may differ because of both the selected training window and the random initialisation of the model parameters.

For each evaluation window, every model produces an overlapping community assignment and a corresponding relative temporal $\mathrm{ONMI}$ score. The final anomaly decision is obtained using one of two approaches. In first, the model scores are aggregated using a running mean, and an anomaly is identified when the resulting score exceeds a $3\sigma$ threshold estimated from the baseline data. In the second approach, majority voting is used, and a window is classified as anomalous when a predefined fraction of the ensemble members individually exceeds the selected threshold. We evaluate both approaches.

\section{Results}
This section first evaluates the proposed model against established results from the literature. We then examine the effect of the diffusion prior across different model architectures through an ablation study. Because DISCO is architecturally most closely related to NOCD and AOCD, we focus primarily on comparisons with these two methods. In the original NOCD study in \cite{shchur2019overlapping}, NOCD variants achieved the best performance on most of the evaluated networks compared to other deep learning models. In AOCD study in \cite{AOCD}, the AOCD further improved the results. Therefore, these models provide the most relevant state-of-the-art baselines for comparison with DISCO and we omit other comparisons.
\subsection{Model Performance}
\subsubsection*{Benchmark}
The baseline models are evaluated under two input configurations: a topology-driven structural-profile (binary adjacency) configuration and an attribute-driven configuration based on node attributes. Tables \ref{tab:graph_topology} and \ref{tab:attribute_driven} report the best baseline configuration for each network together with the corresponding DISCO result. The concatenated attribute--structural configuration, which is supported by DISCO but is not included in the reference comparison, is examined separately in the ablation study in Table \ref{tab:diffusion_ablation}. The reported $\mathrm{ONMI}$ values are presented on a scale from 0 to 100. 100 indicates total agreement, while 0 means there is zero similarity between community assignments.

Across the two input settings, DISCO exceeds the best results reported in \cite{AOCD} in most cases. In the structural-profile input setting (henceforth Structural), DISCO obtains a higher score on five of the six networks. The largest improvements are observed for networks 698 and 414. Network 1684 is the only Structural case in which the reference result is higher, although the difference is marginal. In the node-attribute setting (henceforth Attributes), DISCO also outperforms the reference models on five of the six networks. The largest improvement is obtained similarly for networks 698 and 414. For network 348, the reference GAT result is higher than the best DISCO result.

The role of the prior differs markedly between the two input settings. In the Structural experiments, the best DISCO configuration uses prior for only two of the six networks. Thus, prior-based rewiring does not always improve performance when the input representation already contains binary-adjacency-derived structural information. 
In contrast, every best-performing Attributes-driven DISCO configuration uses the prior. In this setting, the node input representations contain no explicit topological information. The prior introduces graph structure through both the sparse attention mask and the additive bias applied to the attention logits. The results therefore indicate that the prior is particularly beneficial when using attributes and structural information is available.

Finally, in most networks, we also observe higher variability than reported in the reference work. The exact reason for this increased variability is unclear. One possible explanation is the sensitivity introduced by the multi-head attention mechanism and the concatenation of the heads, although the present experiments do not isolate this effect. Interestingly, the results in Table~\ref{tab:diffusion_ablation} provide some indication that the prior may reduce variability, as the standard deviation is smaller in several configurations when the prior is used. However, these differences are generally small and are not present in all networks and input configurations. Improving the stability of the model remains for future work.
\begin{table}[t]

    \centering
    \caption{Comparison of DISCO with the best-performing structural-profile
    and attribute-based models reported in \cite{AOCD} for each Facebook ego network.}
    \begin{subtable}[t]{\textwidth}
        \centering
        \caption{Structural-profile model.}
        \label{tab:graph_topology}

        \begin{tabular}{ccccc}
            \toprule
            \textbf{Graph} &
            \textbf{Best Model type from \cite{AOCD}} &
            \textbf{ONMI} &
            \textbf{Our ONMI} &
            \textbf{Prior} \\
            \midrule
            348  & GCN & $34.6 \pm 1.8$ & \textbf{35.7} $\pm$ \textbf{3.9} & Y \\
            414  & GCN & $55.1 \pm 1.2$ & \textbf{63.0} $\pm$ \textbf{4.3} & N \\
            686  & GCN & $18.7 \pm 0.8$ & $\textbf{19.6} \pm \textbf{1.6}$ & N \\
            698  & GAT & $48.5 \pm 2.6$ & $\textbf{57.1} \pm \textbf{4.2}$ & N \\
            1684 & GAT & $\textbf{40.7}$ $\pm$ $\textbf{2.9}$ & 40.3 $\pm$ 4.2 & N \\
            1912 & GAT & $39.7 \pm 1.3$ & \textbf{41.6} $\pm$ \textbf{4.9} & Y \\
            \bottomrule
        \end{tabular}
    \end{subtable}

    \vspace{1em}

    \begin{subtable}[t]{\textwidth}
        \centering
        \caption{Attribute-based model.}
        \label{tab:attribute_driven}

        \begin{tabular}{ccccc}
            \toprule
            \textbf{Graph} &
            \textbf{Best Model type from \cite{AOCD}} &
            \textbf{ONMI} &
            \textbf{Our ONMI} &
            \textbf{Prior} \\
            \midrule
            348  & GAT & $\textbf{33.6}$ $\pm$ $\textbf{2.6}$ & 32.7 $\pm$ 2.1 & Y \\
            414  & GAT & $53.6 \pm 3.6$ & \textbf{65.7} $\pm$ \textbf{5.2} & Y \\
            686  & GCN & $19.0 \pm 1.4$ & $\textbf{19.1} \pm \textbf{1.5}$ & Y \\
            698  & GAT & $35.3 \pm 4.0$ & $\textbf{60.7} \pm \textbf{5.1}$ & Y \\
            1684 & GAT & $34.6 \pm 2.2$ & \textbf{36.4} $\pm$ \textbf{4.7} & Y \\
            1912 & GCN & $35.9 \pm 3.0$ & \textbf{38.3} $\pm$ \textbf{3.3} & Y \\
            \bottomrule
        \end{tabular}
    \end{subtable}

    \label{tab:aocd_comparison}
\end{table}
\subsubsection*{Ablation Study}

We conduct an ablation study to isolate the contribution of the prior under the three input representations supported by DISCO, i.e., Structural, Attributes, and Concatenated. The structural mode uses rows of the binary adjacency matrix as topology-derived node representations, the Attributes input mode uses only the observed node attributes, and the concatenated mode combines the two representations. Each input type is evaluated both with and without the prior, which results in six configurations per network. The results are reported in Table~\ref{tab:diffusion_ablation}.

\begin{table}[ht]
    \centering
    \caption{Ablation study of the structural prior across different DISCO input configurations. Each input type has a "no prior" and "with prior" comparison, resulting in six configurations per network.}
    \label{tab:diffusion_ablation}
    \begin{tabular}{clcc}
        \toprule
        \textbf{Network} &
        \textbf{Input Mode} &
        \multicolumn{2}{c}{\textbf{ONMI}} \\
        \cmidrule(lr){3-4}
        & &
        \textbf{No Prior} &
        \textbf{With Prior} \\
        \midrule

        \multirow{3}{*}{348}
            & Structural & 34.6 $\pm$ 3.4 & \textbf{35.7} $\pm$ \textbf{3.9} \\
            & Attributes   & 30.2 $\pm$ 3.9 & \textbf{32.7} $\pm$ \textbf{2.1} \\
            & Concatenated       & \textbf{32.3} $\pm$ 4.3 & 31.3 $\pm$ 4.1 \\
        \midrule

        \multirow{3}{*}{414}
            & Structural & \textbf{63.0} $\pm$ \textbf{4.3} & 61.9 $\pm$ 2.5 \\
            & Attributes   & 60.5 $\pm$ 5.9 & \textbf{65.7} $\pm$ \textbf{5.2}\\
            & Concatenated       & \textbf{64.1} $\pm$ 4.0 & 60.3 $\pm$ 4.2 \\
        \midrule

        \multirow{3}{*}{686}
            & Structural & \textbf{19.6} $\pm$ \textbf{1.6} & 19.0 $\pm$ 1.3 \\
            & Attributes   & 19.0 $\pm$ 1.5 & \textbf{19.1} $\pm$ \textbf{1.5}\\
            & Concatenated       & 19.1 $\pm$ 2.2 & \textbf{19.1} $\pm$ 1.9 \\
        \midrule

        \multirow{3}{*}{698}
            & Structural & \textbf{57.1} $\pm$ \textbf{4.2} & 56.9 $\pm$ 4.1 \\
            & Attributes   & 59.1 $\pm$ 6.0 & \textbf{60.7} $\pm$ \textbf{5.1} \\
            & Concatenated       & \textbf{59.3} $\pm$ 5.8 & 58.0 $\pm$ 6.1 \\
        \midrule

        \multirow{3}{*}{1684}
            & Structural & \textbf{40.3} $\pm$ \textbf{4.2} & 36.0 $\pm$ 3.6 \\
            & Attributes   & 35.9 $\pm$ 4.9 & \textbf{36.4} $\pm$ \textbf{4.7}\\
            & Concatenated       & \textbf{41.4} $\pm$ 5.2 & 38.8 $\pm$ 5.4 \\
        \midrule

        \multirow{3}{*}{1912}
            & Structural & 38.1 $\pm$ 4.3 & \textbf{41.6} $\pm$ \textbf{4.9} \\
            & Attributes   & 35.6 $\pm$ 4.7 & \textbf{38.3} $\pm$ \textbf{3.3}\\
            & Concatenated       & 39.6 $\pm$ 3.1 & \textbf{42.5} $\pm$ \textbf{2.9} \\

        \bottomrule
    \end{tabular}
\end{table}

With the six configurations, a prior-enabled model achieves the highest mean $\mathrm{ONMI}$ score on four of the six networks: the Structural model for network 348, the Attribute models for networks 414 and 698, and the Concatenated model for network 1912. A configuration without prior performs best on networks 686 and 1684. The difference between the best prior and non-prior configurations is relatively small for network 686, whereas the advantage of the non-prior configuration is more pronounced for network 1684. The clearest effect of the prior is observed in the Attributes mode, for which adding the prior increases the mean $\mathrm{ONMI}$ score on all six networks. This result is consistent with the design of DISCO: Although the structural information is not included in the node input representation itself, the prior determines which node pairs are available to the sparse attention mechanism and biases their attention logits according to prior-derived proximity. The model in this setting therefore requires indirectly the structural profile via ISM prior.

Directly using structural profile, 
as in Structural mode, does not necessarily improve the $\mathrm{ONMI}$ with prior.
An adjacency graph represents the observed connectivity profile of nodes, whereas the ISM transforms the observed topology into prior probabilities that reflect multi-hop connectivity, shared neighbourhoods, and alternative paths. The resulting structural information is therefore richer and used to determine where information is exchanged and to bias the strength of those interactions, rather than being used directly as node representations. More specifically, using the prior does increase the mean score on networks 348 and 1912 but decreases it on the remaining four networks in Structural mode. 
A plausible interpretation is that using prior provides partly redundant structural information: adjacency-derived information is encoded directly in the node representations, while prior-derived structure simultaneously controls and biases the attention interactions. Depending on the network, these signals may be complementary, but they may also encode similar information or favour completely different relationships. The same observation helps also explain why combining node attributes and structural profiles in Concatenated input mode does not typically produce the highest score. The Concatenated input mode is the best-performing variant for networks 1684 and 1912, whereas a single-input configuration performs best for the other four networks. Furthermore, adding the prior improves the mean ONMI score only for network 1912 in the Concatenated mode. This further suggests that the ISM-derived prior does not provide complementary information when adjacency-derived structural profiles are already included in the input representation.




Because of these observations, we examine how the prior changes the interaction structure of each network in Table~\ref{tab:ism_edge_retention}. As the prior edge ratio is fixed at $r=1.0$, the prior contains the same number of edges as the observed graph; however, the identities of these edges can differ substantially. For five of the six networks, the ISM prior indeed produces considerable rewiring. Only 15--37\% of the original edges are retained for networks 348, 414, 686, 1684, and 1912, corresponding to edge-set Jaccard similarities between 8\% and 22\%. Network 698 is a clear exception: 94\% of its original edges are retained and the edge-set Jaccard similarity is 89\%, which indicates that the prior-derived interaction graph remains highly similar to the observed topology.

These differences confirm that, even at $r=1.0$, the prior 
is typically very different from the original adjacency graph, 
as ISM can substantially reorganise the candidate interactions. Importantly, the magnitude of this rewiring alone does not determine whether prior improves community detection. Network 1684 undergoes the strongest structural change, with only 15\% of its original edges retained, and prior reduces performance in the Structural and 
Concatenated configurations. Network 1912 is also strongly rewired, retaining only 26\% of the original edges, yet prior improves the mean $\mathrm{ONMI}$ score in all three input modes. The benefit of the prior therefore appears to depend on which relationships are preserved, removed, or introduced rather than simply on the extent of the rewiring.

Observations raise the possibility that the individual observed edges are not always the most informative source of community information in these networks. In five of the six cases, the ISM can replace a large fraction of the original edges while retaining, and in some configurations, improving $\mathrm{ONMI}$-score. At the same time, the prior improves every attribute-driven configuration. One possible interpretation is therefore that the node attributes carry a substantial part of the information required to identify the ground-truth communities -- at least in these benchmarking cases -- while topology is just a structural constraint on how those attribute representations interact. From this perspective, the advantage of the prior is 
that it filters and reorganises the observed topology into relationships that are more informative than the raw edges themselves. This could also explain why explicitly adding adjacency profiles to the node representation provides limited or no benefit: raw edge-level information may contain relationships that are not directly informative to the annotated community structure. This interpretation remains network-dependent and partly speculative. For example, the results depend on how the circles are formed and how well they reflect the topology of the graph. The results, at least in these graphs, suggest that using attribute-based node content and diffusion-derived interaction structure may be more effective than representing both attributes and structural profile directly in the node input representations.

\begin{table}[ht]
    \centering
    \caption{Retention and overlap of original Facebook network edges in the ISM-derived structural prior.}
    \label{tab:ism_edge_retention}
    \begin{tabular}{ccccc}
        \toprule
        \textbf{Network} &
        \textbf{Original Edges} &
        \textbf{Original Edges Retained} &
        \textbf{Retained} &
        \textbf{Edge Jaccard} \\
        \midrule
        348  & 3.2K  & 1.1K & 33\% & 20\% \\
        414  & 1.7K  & 0.6K & 34\% & 21\% \\
        686  & 1.7K  & 0.6K & 37\% & 22\% \\
        698  & 0.3K   & 0.3K  & 94\% & 89\% \\
        1684 & 14.0K & 2.0K & 15\% & 8\%  \\
        1912 & 30.0K & 7.8K & 26\% & 15\% \\
        \bottomrule
    \end{tabular}
\end{table}

\section{Cybersecurity Use Case: Dynamic Community Detection for Network Anomaly Detection}
\label{sec:anomaly_detection}

In this experiment, the underlying assumption is that communication in a stable network environment exhibits recurring structural patterns. Devices that regularly communicate with one another are expected to form relatively stable, although potentially overlapping, communities. An event that changes the normal communication behavior---for example, the disappearance of an established connection---may also alter these community assignments. We therefore hypothesise that rapid changes in the inferred overlapping community structure can serve as an indicator of anomalous network behavior. On the other hand, longer 
persistent changes in patterns 
can be interpreted as a signature of baseline drift, a structural fingerprint that reveals when the network departs from its normal operating regime.


\subsection{Evaluation and Interpretation}

\begin{table}[htbp]
    \centering
    \caption{Detection performance across Anomaly Scenarios using the
    \(3\sigma\) threshold and majority-voting methods.}
    \label{tab:PerformanceTestCases}

    \renewcommand{\arraystretch}{1.25}
    \setlength{\tabcolsep}{5pt}
    \rowcolors{2}{gray!8}{white}

    \begin{tabularx}{\linewidth}{
        c
        >{\raggedright\arraybackslash}X
        *{2}{S[table-format=1.4]}
        S[table-format=2.1]
        *{2}{S[table-format=1.4]}
        S[table-format=2.1]
    }
        \toprule
        &
        & \multicolumn{3}{c}{\textbf{\(3\sigma\) Threshold}}
        & \multicolumn{3}{c}{\textbf{Majority Voting}} \\
        \cmidrule(lr){3-5}
        \cmidrule(lr){6-8}

        \textbf{\#}
        & \textbf{Anomaly Scenario}
        & {\textbf{F1}}
        & {\textbf{FPR}}
        & {\makecell{\textbf{Latency}\\\textbf{(min)}}}
        & {\textbf{F1}}
        & {\textbf{FPR}}
        & {\makecell{\textbf{Latency}\\\textbf{(min)}}} \\
        \midrule

        1. & Stale Asset
        & {0.98} & {$10^{-3}$} & {0.7}
        & {0.98} & {$10^{-3}$} & {0.6} \\

        2. & Changed Device Functionality
        & {0.16} & {$10^{-4}$} & {1.3}
        & {0.92} & {$10^{-2}$} & {1.7} \\

        3. & Fingerprinted Traffic
        & {0.64} & {$10^{-3}$} & {1.9}
        & {0.43} & {$10^{-3}$} & {1.9} \\

        4. & New Port Between Hosts
        & {0.02} & {$10^{-3}$} & {16.1}
        & {0.00} & {$10^{-5}$} & {N/A} \\

        5. & Changed Communication Frequency
        & {0.11} & {$10^{-3}$} & {41}
        & {0.12} & {$10^{-2}$} & {27} \\

        6. & Missing Communication
        & {0.96} & {$10^{-2}$} & {0.6}
        & {0.64} & {$10^{-2}$} & {1.3} \\

        7. & Pattern Repeat Failure
        & {0.62} & {$10^{-3}$} & {0.7}
        & {0.45} & {$10^{-3}$} & {0.8} \\

        8. & Pattern Stops Repeating
        & {0.92} & {$10^{-2}$} & {0.6}
        & {0.54} & {$10^{-3}$} & {2.0} \\

        9. & More Traffic Between Known Hosts
        & {0.09} & {$10^{-2}$} & {6.3}
        & {0.05} & {$10^{-2}$} & {22} \\

        10. & Less Traffic Between Known Hosts
        & {0.63} & {$10^{-3}$} & {0.9}
        & {0.71} & {$10^{-3}$} & {0.8} \\
        \bottomrule
    \end{tabularx}
\end{table}

The detection performance across the Anomaly Scenarios is summarised in Table \ref{tab:PerformanceTestCases}. For operational anomaly-detection systems, two properties are particularly important: a low false-positive rate (FPR), defined as the proportion of normal observations incorrectly classified as anomalous, and a short detection latency, defined as the elapsed time between the onset of an anomaly and the first correct alert. A low FPR is essential because excessive false alarms increase the workload of system operators and may reduce confidence in the detection system, whereas short latency enables a timely response to abnormal conditions. We additionally report the F1-score, the harmonic mean of precision and recall, as it is a standard performance metric in machine-learning studies. However, as discussed below, and in \cite{koistinen2026spatiotemporalattentiongraphneural}, F1-scores should be interpreted cautiously in anomaly detection systems.

Results are reported for two decision strategies. In the first, model-specific detection thresholds are derived from the baseline period using a 3$\sigma$ criterion. In the second, the binary decisions of the ensemble members are aggregated via majority voting. The voting threshold was set to 0.6, requiring more than half of the models to agree before a time window was classified as anomalous. The appropriate threshold may depend on the variability and stability of the underlying communication network. In the present setting, the ensemble outputs were sufficiently stable during the baseline period. Accordingly, a majority-voting threshold of greater than 50\% agreement was adopted after vetting the baseline votes across the Anomaly Scenarios.

\begin{figure}[ht]
    \centering
    \vbox{%
        \hbox{\includegraphics[width=0.95\linewidth]{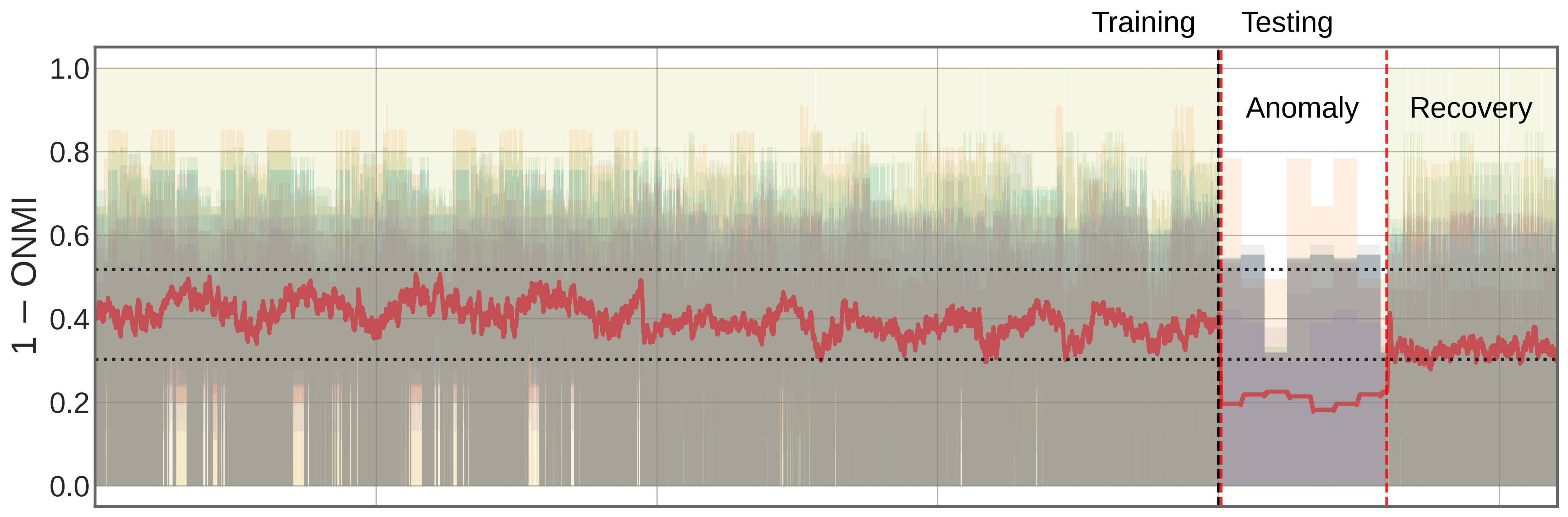}}%
        \nointerlineskip
        \hbox{\includegraphics[width=0.95\linewidth]{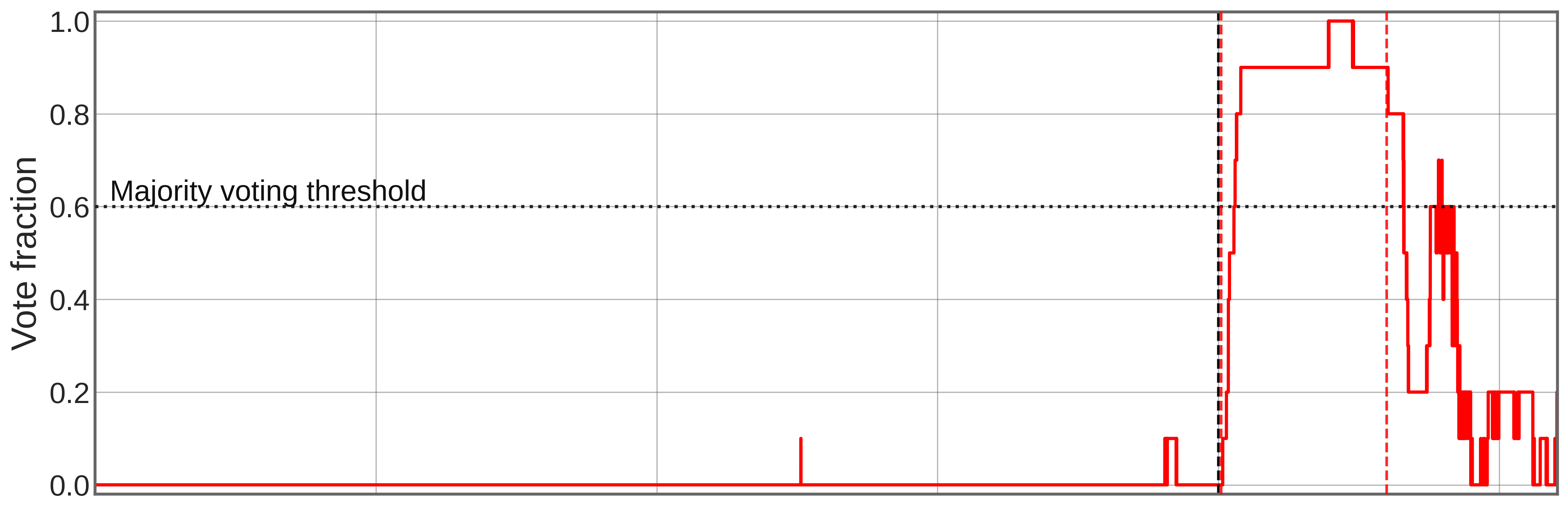}}%
        \nointerlineskip
        \hbox{\includegraphics[width=0.95\linewidth]{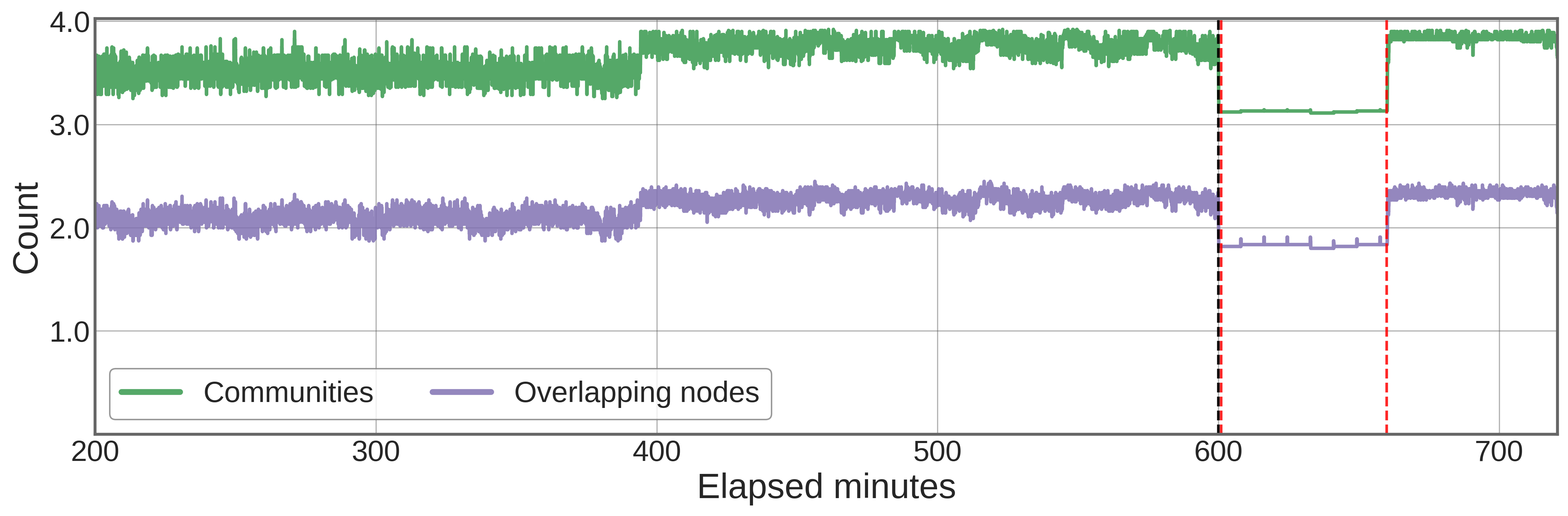}}%
    }
    \caption{Time-series summary of Anomaly Scenario 1, using a 30-second rolling mean. Top: relative community-assignment distance ($1-\mathrm{ONMI}$), showing individual model trajectories (transparent lines), the aggregate trajectory (solid red), and decision thresholds (dotted horisontal lines). Middle: fraction of models voting for an anomaly, with the majority-vote threshold indicated by the dotted line. Bottom: smoothed overlapping-node and community counts. The black dashed line marks the transition from training to testing, and the red dashed lines delimit the anomaly interval; the subsequent period shows recovery phase.}
    \label{fig:thresholdtimeseries}
\end{figure}

Figure \ref{fig:thresholdtimeseries} illustrates the temporal evolution of the relative $\mathrm{ONMI}$-based anomaly score in Anomaly Scenario 1. The rolling-mean trajectory in the upper panel reveals a clear structural change when a device disappears from the communication network, while the middle panel shows the majority-voting results across the experiment. Following the removal of the device, agreement among the inferred community assignments becomes more stable, while both the mean number of overlapping nodes and the number of detected communities decrease, as seen in the bottom panel. This behaviour suggests that interrupting the communication of a device causes the remaining communities to become more clearly separated. However, this is not always the case: the community and overlapping-node counts do not necessarily change during an anomaly, even when the relative $\mathrm{ONMI}$ exceeds the detection thresholds.

During the subsequent recovery phase, the relative $\mathrm{ONMI}$ and the community and node counts do not fully return to their original baseline levels. Instead, a small but persistent baseline shift remains observable, with several models continuing to vote for an anomaly after normal communication has resumed. This increases the FPR and suggests that the detector may need recalibrating the thresholds, or allowing for a longer recovery phase.

Several additional observations can be made from the results. Many cases exhibit relatively high FPR, which is largely attributable to the recovery phase. In almost all cases, recovery is noisy. For example, in Figure \ref{fig:thresholdtimeseries}, the running-mean trajectory shows a shifted baseline during recovery and continues to trigger alerts, thereby increasing the FPR. A similar effect is visible in the post-anomaly phase of the majority-voting plot, where some ensemble members continue to produce alerts after the main anomalous event. In addition to increasing the FPR, a shifted baseline or, in this case, incomplete recovery reduces the F1-score. A low F1-score therefore does not necessarily imply that a detector is operationally ineffective. In the present application, a detector may identify the onset of an anomaly rapidly and with very few false positives in recovery, yet stop producing alerts as it adapts to the altered network state. This is a known property of relative scoring: the anomaly phase becomes the new normal. Such behaviour results in missed anomalous time windows, reducing recall and consequently the F1-score. From an operational perspective, however, detecting every anomalous time point may be less important than providing an early and reliable indication that an abnormal event has occurred.

For example, in Anomaly Scenario 2, the $3\sigma$ strategy achieves an F1-score of only 0.16, but an FPR of 
$10^{-4}$ and a detection latency of 1.3 minutes. This indicates that the method is sensitive to the onset of the anomaly and then misses most of the subsequent anomaly signals. During the recovery phase, in turn, the model produces very few false positives. By comparison, majority voting achieves a substantially higher F1-score of 0.92, but also a higher FPR of 0.05 and a longer latency of 1.7 minutes. These results demonstrate that F1-score, FPR, and detection latency should be considered jointly, rather than relying on the F1-score as the sole criterion of detector performance.

Finally, although the proposed method produces a global anomaly score based on the change in relative $\mathrm{ONMI}$, it also enables the underlying causes of the detected anomaly to be identified. In particular, the reduction in temporal community similarity can be decomposed into node-level contributions, revealing which devices changed their community memberships, lost established affiliations, formed unexpected relationships, or otherwise contributed most strongly to the anomaly. Figure \ref{fig:placeholder} presents the normalised device-level contributions separately for each ensemble member in the same Anomaly Scenario during anomaly. Most models (3,4,5,6,8,9,10) produce consistent explanations and identify the disappeared device as the principal contributor to the altered community assignments. 
The decomposition can also be extended to examine associated changes in edges or node features. Thus, the community-based formulation is not limited to indicating that an anomaly has occurred, but also provides interpretable evidence about its likely source, supporting subsequent investigation of the detected event. 

\begin{figure}
    \centering
    \includegraphics[width=0.75\linewidth]{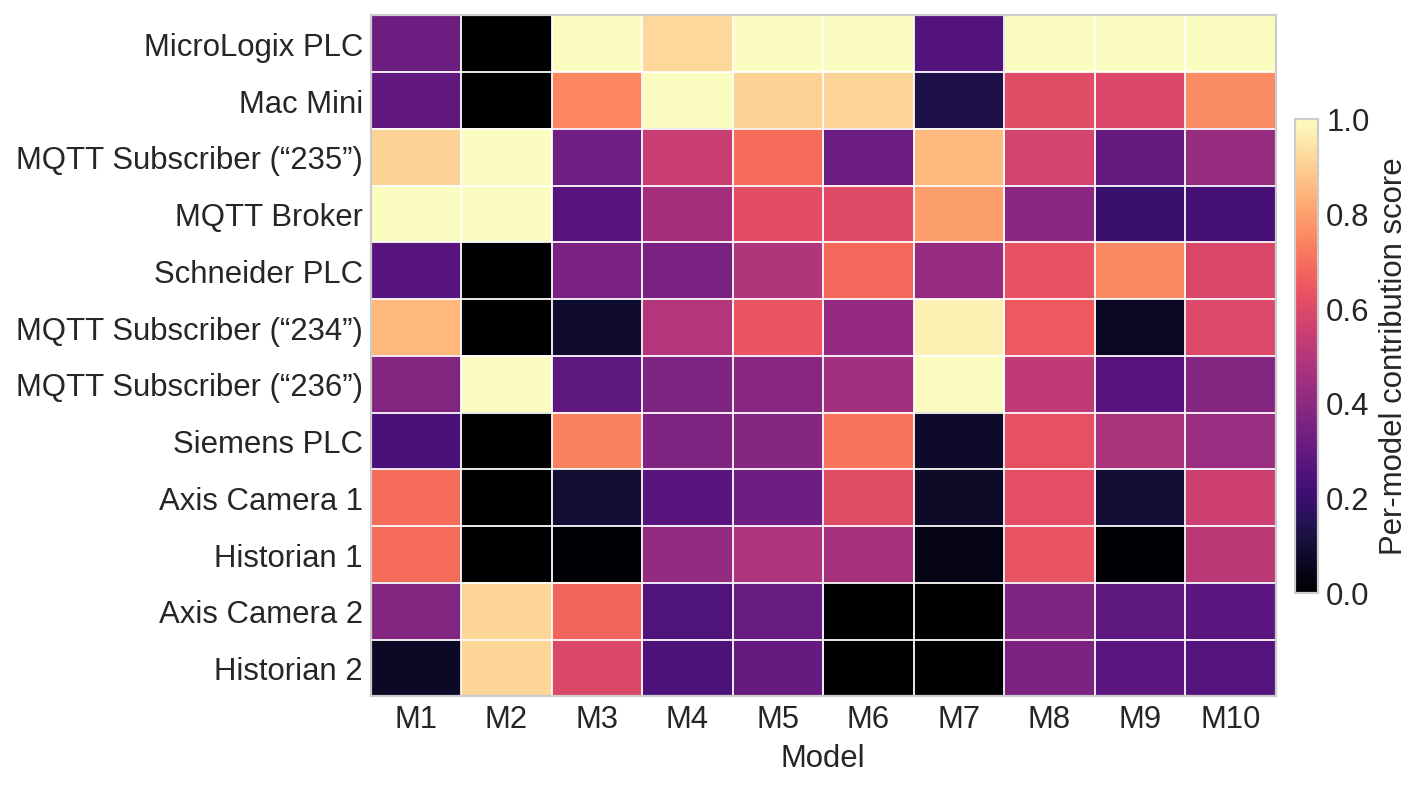}
    \caption{Device importance normalised model by model in anomaly phase in Anomaly Scenario 1. Higher score means higher relative change in $\mathrm{ONMI}$ score. Most models detect the correct devices responsible for anomaly.}
    \label{fig:placeholder}
\end{figure}

\subsection{Limitations and Opportunities}

The present experiment has several limitations. First, it considers only a few controlled anomaly types and should therefore be regarded as a demonstration rather than a comprehensive anomaly-detection benchmark. Second, temporal community changes may also be caused by legitimate operational events, including maintenance, changes in workload, device restarts, or process transitions. A low $\mathrm{ONMI}$ score should consequently be interpreted as evidence of structural change or a baseline shift, rather than as definitive evidence of an attack. Third, the method depends on several design choices, including the graph construction procedure, aggregation-window length, model training windows, ensemble aggregation rule, and anomaly-thresholding strategy. The sensitivity of the results to these choices should be examined in future work. Fourth, DISCO is transductive, meaning that previously unseen devices cannot be directly included without retraining the model. In practice, DISCO would therefore need to be retrained when new devices are added to the network. We do not consider this a major limitation in the present OT setting, where the device population is typically relatively stable and changes to the network are generally less frequent than in highly dynamic IT environments. Retraining could therefore be executed when the network configuration changes. Finally, the use of degree and PageRank as pseudo-features provides only a limited description of device behaviour. Future extensions could incorporate protocol information, device roles, flow statistics, temporal features, or learned representations. For example, anomaly 4 could plausibly be detected, if port-related attributes were introduced to the graph.

Despite these limitations, the experiment illustrates the potential of overlapping community detection as an interpretable basis for anomaly detection. The results show that deviations from stable community assignments can reveal changes in complex communication environments and that the affected nodes and relationships can be inspected to help explain the resulting anomaly signal. The temporal trajectories presented in the figures also provide an intuitive representation of the network baseline and its variability or stability over time. Such visualisations may provide a useful tool for operational experts to assess network stability, identify baseline shifts, and investigate the structural changes underlying detected anomalies.

\section{Concluding Remarks}

This work introduced Diffusion-Induced Spatial Attention Community Detection (DISCO), a neural model for overlapping community detection that combines diffusion-derived structural information, sparse multi-head attention, and non-negative community-affiliation learning. The diffusion prior allows nodes to interact with structurally relevant nonlocal nodes rather than being restricted to observed immediate neighbours. By sparsifying these interactions, DISCO provides a compromise between the limited receptive field of conventional graph neural networks and the quadratic cost of unrestricted global attention.

Benchmarking on six Facebook ego networks showed that DISCO was competitive with established GCN- and GAT-based methods. It outperformed the best reported reference results in most of the topology- and attribute-driven comparisons. The clearest benefits were observed in the attribute-driven setting, where the prior supplied structural information that was not included in the node inputs. However, prior was not universally beneficial. When full adjacency-derived structural information was already included in the input representation, the effect of the prior varied accross the networks. Diffusion-based rewiring may introduce useful nonlocal interactions, but it may also replace informative local edges or duplicate structural information already available to the model. Similarly, directly combining node attributes and structural profiles did not consistently produce the best results. The prior parameters and input representation should therefore be selected according to the characteristics of the network rather than treated as universally optimal design choices.

Our cybersecurity demonstration illustrated how changes in overlapping community assignments can be used as an interpretable anomaly signal or as baseline monitoring in dynamic communication networks. By comparing the similarity of community assignments between consecutive time windows, relative community assignment comparison made it possible to identify periods in which the communication structure changed substantially. Node-level contributions further helped identify which devices were primarily responsible for the detected changes. Nevertheless, this experiment should be regarded as a proof of concept rather than a complete anomaly-detection system.

Future work should investigate adaptive methods for selecting prior parameters and edges, as well as more selective mechanisms for combining attributes and topology. Evaluation should also be extended to larger and more diverse networks and compared with recent graph-transformer and diffusion-based community-detection methods. For the cybersecurity application, an important next step is to develop an inductive temporal model that can update community assignments online, distinguish abrupt anomalies from gradual baseline drift, and use richer device and communication features.

Overall, the results show that prior-guided attention is a promising approach for using higher-order graph structure in overlapping community detection. Its main benefit is not that prior always enhances community recovery with every network, but that it provides a principled and controllable mechanism for extending the graph learning beyond observed local edges.

\newpage

\section{APPENDIX A: Influence-Spreading Model Weights}

The influence-spreading process is governed by several parameters, including maximum length $L$, spreading time $T$, and the weights $w$ that are assigned to the existing links. In our work, to construct the structural prior, we used a uniform weight of ($w=0.05$) to all links. This choice is motivated both by previous studies of the influence-spreading model \cite{koistinen2025importance,kuikka2018influence,kuikka2024detailed} and by the need to preserve informative variation in the resulting spreading probabilities. If the link weights are too large, spreading becomes almost certain and the pairwise probabilities approach one. The prior then loses its ability to distinguish between strongly and weakly related node pairs. Conversely, if the weights are too small, influence does not propagate beyond the immediate neighbourhood, and the resulting probabilities fail to reveal the community structure. An intermediate weight therefore allows the spreading probabilities to reflect cohesive regions of the graph without becoming either saturated or negligible.

\begin{figure}[ht]
    \centering
    \includegraphics[width=0.95\linewidth]{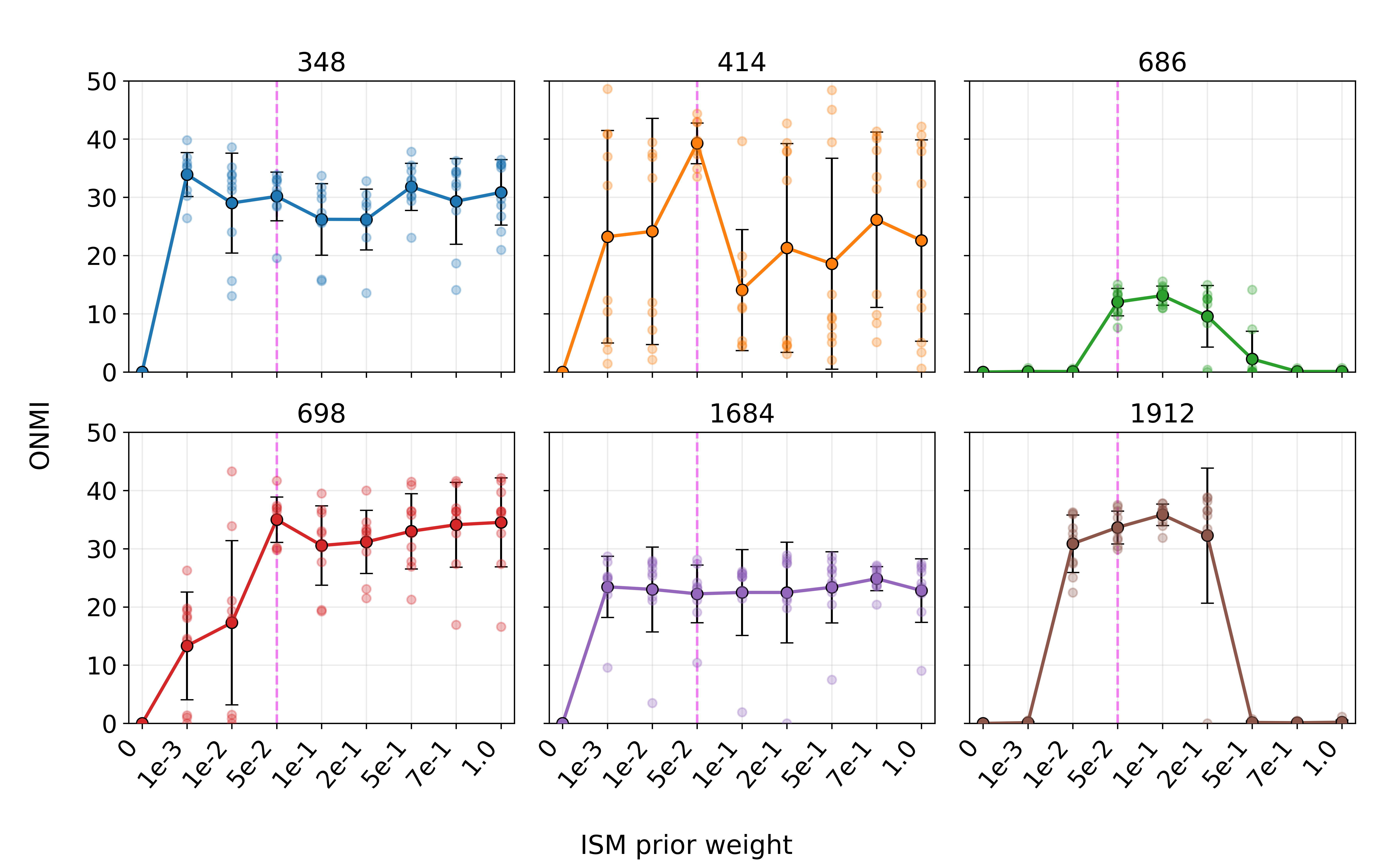}
    \caption{Weight impact for Facebook networks with different weights in dense setting.}
    \label{fig:appendixa1}
\end{figure}

Figure \ref{fig:appendixa1} illustrates this effect for Facebook networks in the dense setting in structural-profile input mode, in which the complete ISM prior graph is supplied to the model without thresholding it. The results indicate that ($w=0.05$) provides smallest variation in most cases, and achieves strong $\mathrm{ONMI}$-score performance across 20 repeated runs. Importantly, this value was selected \textit{a priori} rather than chosen \textit{post-hoc} to optimise the reported results. However, we remind that the appropriate scale of the weights necessarily depends on the size, density, and connectivity patterns of the network. The selected value of weight is therefore based on both earlier findings and the expected spreading dynamics of the model in these particular small social networks.

\begin{figure}[ht]
    \centering
    \includegraphics[width=0.95\linewidth]{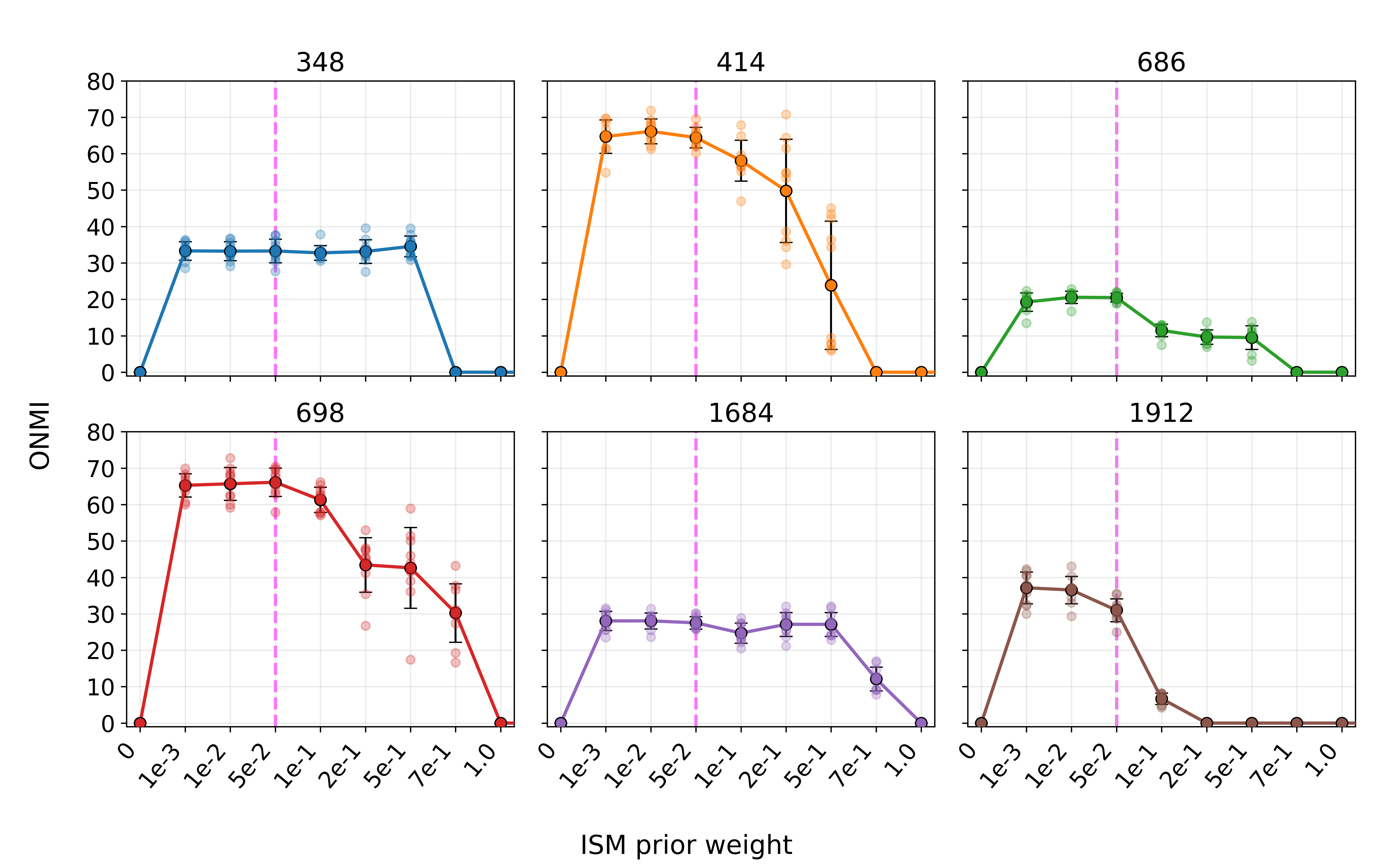}
    \caption{Weight impact for FB networks in filtered case. Edge ratio $r=1.0$, 20 runs per weight per graph. }
    \label{fig:appendixa2}
\end{figure}

We further examine the effect of the link weights in a more practical sparse setting, since dense all-pairs attention becomes computationally infeasible for large networks. The prior is sparsified by thresholding the ISM probabilities so that an edge ratio $r=1.0$ is retained. This approach reduces the number of attention interactions and, as the results show in Figure \ref{fig:appendixa2}, can improve community-detection performance while requiring much less computation than the dense setting. 
In this setting, the prior contains the same number of edges as the original adjacency graph, although the retained edge set is not identical to the original one. The results show that ($w=0.05$) does not necessarily produce the highest $\mathrm{ONMI}$ for every network. Instead, it is a compromise that yields competitive performance across networks. This variation shows that the prior should be assessed carefully and may require network-specific calibration.

When $(w=0)$, the spreading probabilities contain no information about the network topology. In other words, the graph does not have edges from which to infer communities. However, even small positive weights are sufficient to recover meaningful structural patterns. The results reflect to a structure when using solely an adjacency graph as a prior. The choice of weight is therefore not very sensitive to the results when using comparatively low weights. As the weight increases, longer spreading paths become more probable and the contrast between structurally important and unimportant node pairs decreases. At larger values, such as $w=0.1$, less informative node pairs may consequently enter the retained prior graph and displace more informative relationships. The core edges within dense groups start to dominate, while bridging edges start to diminish. Finally, when weights are close to $1$, the corresponding edge probabilities are close to 1, and prior becomes a binary adjacency graph, which, again, is uninformative. The graph has an altered structure, and without controlling the edge ratio, would become a complete graph. In thresholded case, the survived nodes probabilities close 1, and it will therefore become a binary adjacency graph, that has different structure than the original graph. Furthermore, as weights are increased, less central nodes start losing connections against high centrality nodes. This results in breaking down of components into smaller, which causes fragmentation of the distant nodes as they lose their only informative edges. This is considered a limitation. One possible safeguard would be to require that every node retain at least one incident edge when the prior graph is thresholded. Such a constraint could reduce the risk of isolating peripheral nodes, but its effect on community-detection performance is included in our future work. 

\clearpage

\section{APPENDIX B: Effect of the Prior Edge Ratio}

The sparsity of the diffusion-derived prior can be controlled either by applying a fixed probability threshold or by conveniently specifying an edge ratio. We define the edge ratio as
\[
r = \frac{|\mathcal{E}{\mathrm{prior}}|}{|\mathcal{E}|},
\]
where ($\mathcal{E}_{\mathrm{prior}}$) is the retained prior edge set and ($\mathcal{E}$) is the observed edge set. Thus, ($r=1.0$) means that the prior graph contains the same number of edges as the original graph. Importantly, this does not imply that the two graphs contain the same node pairs. The prior edges are selected according to their influence-spreading probabilities and may therefore replace observed edges with structurally relevant nonlocal interactions. This is shown in Table \ref{tab:ism_edge_retention}.

Controlling the edge ratio has two purposes. First, it provides a direct mechanism for balancing structural expressiveness and computational cost. Increasing $r$ exposes the attention mechanism to a larger set of candidate interactions. This includes multi-hop connections. Decreasing $r$, in contrast, removes low-scoring interactions and produces a sparser attention graph. Second, the edge ratio can affect the quality of the detected community structure. A sparse prior may suppress noisy or weakly informative relationships, whereas a denser prior may reveal higher-order dependencies. These might not been present in the original adjacency graph.

\begin{figure}[ht]
    \centering
    \includegraphics[width=0.95\linewidth]{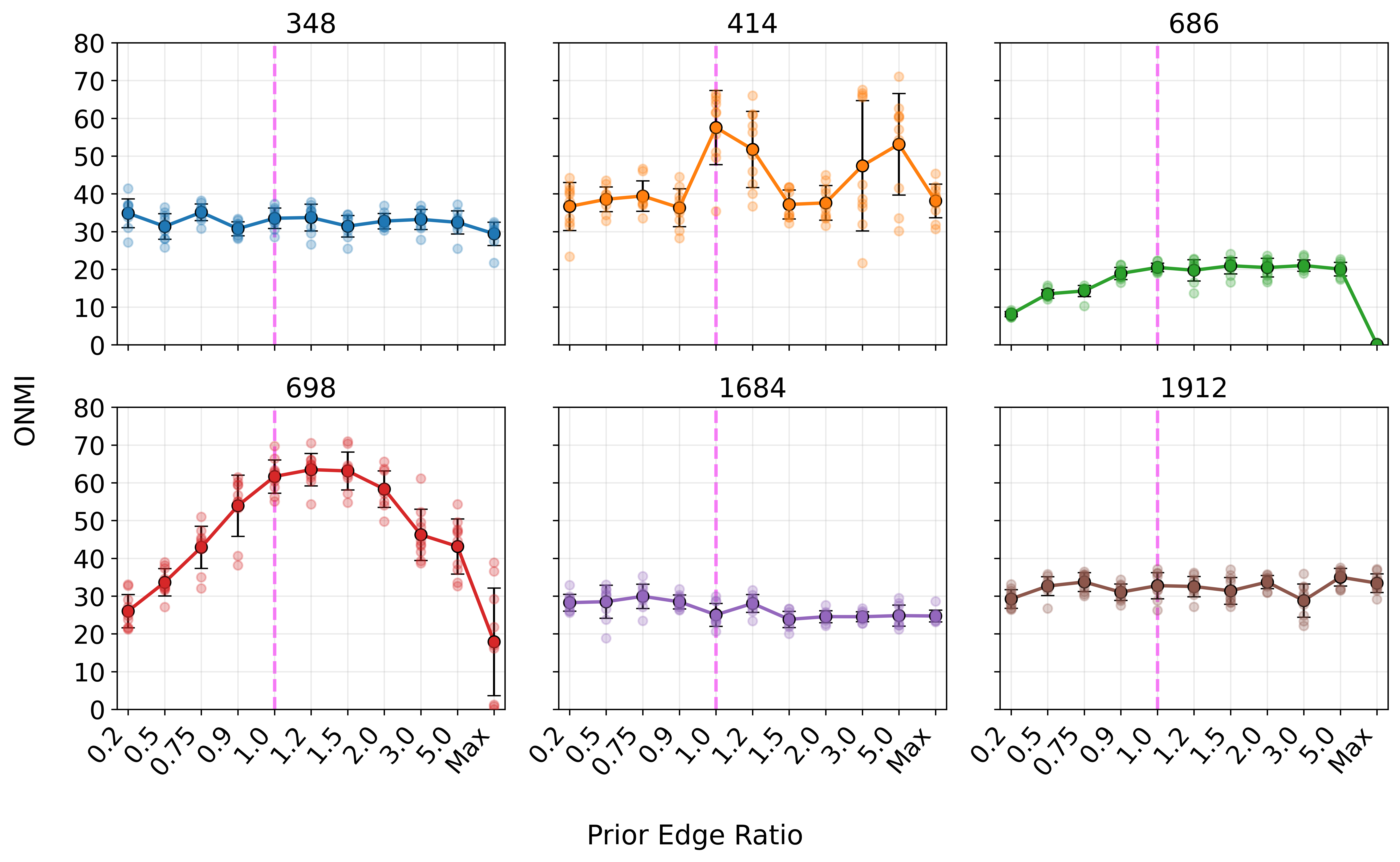}
    \caption{The $\mathrm{ONMI}$ accross edge ratios. With pink dashed line, we show the weight used in benchmarking. The Max indicates for maximum amount of edges possible for each graph. For most graphs, the maximum edge ratio is between 6 and 10; For 1684, it is 24. With max weight, the results are similar to dashed line region in Figure \ref{fig:appendixa1}.}
    \label{fig:appendixb1}
\end{figure}

Figure~\ref{fig:appendixb1} reports the $\mathrm{ONMI}$ score as a function of the edge ratio for the Facebook ego networks in structural-profile input mode. The results show that increasing the number of prior edges does not always improve performance. For networks 348, 686, and 1684, the highest observed $\mathrm{ONMI}$ scores are obtained with edge ratios below $1.0$. This suggests that, in these networks, the prior can act as a structural filter: retaining only the most strongly supported relationships may remove interactions that obscure the underlying community structure. It also indicates that the relevant diffusion information is concentrated in a relatively small subset of node pairs and that adding further edges may introduce redundant or weakly informative interactions.

Networks 698 and 414 exhibit a different pattern, as their performance can benefit from edge ratios $\geq1.0$. One possible explanation is that their observed adjacency structures do not contain all relationships required to distinguish the overlapping communities. In such cases, diffusion-induced nonlocal edges may connect nodes that are structurally related through multiple paths despite not being directly adjacent. These additional interactions enlarge the receptive field of the attention mechanism and may make higher-order community structure easier to identify.

The results do not support a universally optimal edge ratio. Instead, the preferred level of sparsity appears to depend on graph properties such as density, connectivity, noise, and the distribution of overlapping memberships. Nevertheless, the comparatively strong performance obtained at $r=1.0$ across the evaluated networks supports its use as a practical default that neither increases the number of attention interactions nor requires network-specific tuning. Ratios below $1.0$ can reduce both computation and memory consumption, whereas ratios above $1.0$ trade additional computational cost for a broader interaction space. Selecting the edge ratio can thus be viewed as a model-selection problem that is a trade-off between predictive performance and computational efficiency.

\clearpage

\section{APPENDIX C: Network Environment Devices}\label{app:device_list}
All equipment resides on a single IPv4 \texttt{/24} subnet (\texttt{10.100.0.0/24}). Devices and their roles are listed in Table~\ref{tab:device_list}.
\begin{table}[h]
\small
\centering
\caption{Network devices and their roles.}
\label{tab:device_list}
\begin{tabular}{p{.2\columnwidth} p{.17\columnwidth} p{.17\columnwidth} p{.35\columnwidth}}
\toprule
\textbf{Device}                        & \textbf{Role}                     & \textbf{IP Address / MAC}   & \textbf{Notes} \\
\midrule
Mac Mini (Mac OS X)                    & HMI/Webinterface host             & \texttt{10.100.0.108}\newline\texttt{14:9d:99:82:88:da} & Interacts with Siemens web UI to generate HTTP traffic as well as the MicroLogix PLC.                                       \\
\midrule
Siemens S7-1200 PLC                    & Master PLC (Modbus-TCP server)    & \texttt{10.100.0.204}\newline\texttt{28:63:36:86:9f:55} & Holds the central data-block (DB1) accessed by other PLCs; also hosts web server interface with Boolean controls.           \\
\midrule
Schneider M221 PLC                     & Slave PLC (Modbus-TCP client)     & \texttt{10.100.0.202}\newline\texttt{00:80:f4:0e:ce:24} & Receives register write commands from Siemens PLC; sends Modbus-TCP packets to Historian \#1.                               \\
\midrule
MicroLogix 1100 PLC                    & Secondary PLC (CIP/Modbus)        & \texttt{10.100.0.201}\newline\texttt{f4:54:33:9d:1d:1b} & Communicates with Historian \#2 and Mac Mini via CIP protocols.                                                             \\
\midrule
Linux Mint OS \#2 VM (Historian \#1)   & Data historian and MQTT Publisher & \texttt{10.100.0.231}\newline\texttt{bc:24:11:8e:57:1c} & Stores Schneider PLC \& Camera \#1 data; forwards data to MQTT Broker on topic \mbox{``plc/schneider''}.                    \\
\midrule
Linux Mint OS \#3 VM (Historian \#2)   & Data historian and MQTT Publisher & \texttt{10.100.0.232}\newline\texttt{bc:24:11:47:2d:1c} & Stores MicroLogix PLC \& Camera \#2 data; forwards data to MQTT Broker on topic \mbox{``plc/micrologix''}.                  \\
\midrule
Linux Mint OS \#4 VM (MQTT Broker)     & MQTT Broker                       & \texttt{10.100.0.233}\newline\texttt{bc:24:11:f3:4e:79} & Publishes values received from Historian \#1 and Historian \#2 to subscribed hosts (OS \#5, \#6, \#7) on respective topics. \\
\midrule
Camera \#1: Axis Comm.                 & Video stream source               & \texttt{10.100.0.128}\newline\texttt{ac:cc:8e:3f:0e:a3} & Provides HTTP video feed \& telemetry.                                                                                      \\
\midrule
Camera \#2: Axis Comm.                 & Video stream source               & \texttt{10.100.0.129}\newline\texttt{ac:cc:8e:33:b5:67} & Provides HTTP video feed \& telemetry.                                                                                      \\
\midrule
Linux Mint OS \#5 VM (MQTT Subscriber) & MQTT Subscriber                   & \texttt{10.100.0.234}\newline\texttt{bc:24:11:27:84:df} & Subscriber to \mbox{``plc/schneider''}.                                                                                     \\
\midrule
Linux Mint OS \#6 VM (MQTT Subscriber) & MQTT Subscriber                   & \texttt{10.100.0.235}\newline\texttt{bc:24:11:42:6d:b3} & Subscribes to generic topic \mbox{`plc/\#'}; receives Schneider and MicroLogix PLC data values.                             \\
\midrule
Linux Mint OS \#7 VM (MQTT Subscriber) & MQTT Subscriber                   & \texttt{10.100.0.236}\newline\texttt{bc:24:11:72:c9:16} & Subscribes to topic \mbox{`plc/micrologix'}; receives MicroLogix PLC data values.                                           \\
\bottomrule
\end{tabular}

\end{table}

\clearpage

\bibliographystyle{unsrt}
\bibliography{references}  


\end{document}